\documentclass[preprint]{aastex}
\input psfig.sty

\def\degree{$^{\circ}$}

\def\Volker{V$\ddot{\textrm{o}}$lker$\;$}

\def\tlabel#1{\textit{#1}}
\def\vkm{km s$^{-1}$}

\def\nam{newly swept-up}

\def\cm3{cm$^{-3}$}
\slugcomment{\today}

\begin{document}
\title{Hydrodynamical Simulations of Jet- and Wind-driven
Protostellar Outflows}

\author{Chin-Fei Lee, James M. Stone, Eve C. Ostriker, and Lee G. Mundy}
\affil{Astronomy Department, University of Maryland,
    College Park, MD 20742}
\email{chinfei@astro.umd.edu, jstone@astro.umd.edu, ostriker@astro.umd.edu, lgm@astro.umd.edu}

\begin{abstract}
We present two-dimensional hydrodynamical simulations of
both jet- and wind-driven models for
protostellar outflows in order to make detailed comparisons
to the kinematics of observed molecular outflows.
The simulations are performed with the ZEUS-2D hydrodynamical code
using a simplified equation of state, simplified cooling and 
no external heating, and no self-gravity.

In simulations of steady jets, swept-up ambient gas forms a thin shell that
can be identified as a molecular outflow. We find a simple ballistic
bow-shock model is able to reproduce the structure and transverse velocity
of the shell.
Position-velocity (PV) diagrams for the shell cut along the outflow axis
show a convex spur structure with the highest velocity at the bow tip,
and low-velocity red and blue components at any viewing
angle.
The power-law index of the mass-velocity relationship ranges
from 1.5 to 3.5, depending strongly on the inclination. 
If the jet is time-variable, 
the PV diagrams show multiple convex spur structures 
and the power-law index becomes smaller than the steady jet
simulation.

In simulations of isothermal steady wide-angle winds, swept-up ambient gas
forms a thin shell which at {\em early} stages has a similar shape to the shell in
the jet-driven model; it becomes broader at later times.
We find the structure and kinematics of the
shell is well described by
a momentum-conserving model similar to that of \citet{Shu1991}.
In contrast to the results from jet simulations, the PV diagrams for the shell
cut along the outflow axis show a lobe
structure tilted with source inclination, 
with components that are primarily either red or blue unless the
inclination is nearly in the plane of sky.
The power-law index of the mass-velocity
relationship ranges from 1.3 to 1.8.
If the wind is time-variable,
the PV diagrams also show multiple structures, and the power-law index
becomes smaller than the steady wind simulation.

Comparing the different simulations with observations, we find
that some outflows, e.g., HH 212,
show features consistent with the jet-driven model, while others, 
e.g., VLA 05487, are consistent with the wind-driven model. 
\end{abstract}


\section{Introduction}

Protostellar jets and molecular outflows are often
associated with the same young stellar objects
\citep{Lada1985,Bachiller1996}.
They provide unique information about the mass-loss properties
of young stars and give insight into the star forming process itself.
However, the physics connecting the two phenomena remains unclear.

Protostellar jets are generally detected as optical jets in [SII]
$\lambda\lambda$6716, 6731 and H$\alpha$ \citep{Ray1996,Reipurth1999},
as infrared jets in H$_2$ \citep{Coppin1998,Zinnecker1998}, or
as radio jets in the radio continuum \citep{Anglada1995}.
They are also observed in some molecules, e.g.,
SiO \cite[L 1448,][]{Dutrey1997} and CO \cite[HH 211,][]{Gueth1999}. Generally,
the jets are highly collimated structures, consisting of a series 
of knots and bow-shocks, extending as far as a few pc out from the
central stars \cite[e.g., HH 111,][]{Reipurth1997}. 

Molecular outflows are generally seen  in CO and some other molecules,
as bipolar quasi-conical or -parabolic
structures surrounding the jet.
The observational properties of molecular outflows are summarized in
\citet{Masson1993}, \citet{Bachiller1996}, 
\citet{Cabrit1997}, and \citet{Richer2000}.
Molecular outflows are much more massive and normally
less well collimated than the jets.
They have a simple power-law mass
distribution with velocity 
\citep{Masson1992,Stahler1994}.
In many cases, the velocities in the molecular outflows increase with 
distance from the source.  At low velocities,  the outflow appears as a
limb-brightened shell surrounding a cavity, while at high velocities
the outflow sometimes becomes more jet-like. 
There is also evidence that
outflow velocities are directed primarily along the major axis of the
outflow, with little overlap of blueshifted and redshifted emission within
the same flow lobe \citep{Meyers1991,Lada1996}.
In a small sample of outflows known to be associated with jets,
position-velocity (PV) diagrams constructed in a cut along the 
outflow axis reveal two different kinematic structures:
a parabolic structure originating at the driving source,
and a convex structure with the high velocity tip
near the H$_2$ bow shock structure \citep{Lee2000}.

A number of models have been proposed to explain the connection between
protostellar jets and molecular outflows.
Currently, models in which the outflows are driven either by a highly
collimated jet 
\citep{Raga1993,Masson1993,Chernin1994}
or by a wider-angle wind
\citep{Shu1991,Li1996b,Shu2000}
are the most popular.
In the former, the bow shock originating
at the head of the jet sweeps the ambient material into a thin shell which
might be identified as the molecular outflow.
A number of simulations
have been carried out to investigate the hydrodynamics of  protostellar jets 
\citep{Blondin1989,Blondin1990,Stone1993a,Stone1993b,
Stone1994,Biro1994,Suttner1997,Smith1997,Downes1999}.
Such simulations show that steady, collimated jets are 
able to produce qualitatively
the structure and kinematics of some  molecular outflows, e.g., NGC 2264G
\citep{Smith1997}.
In the wind-driven shell
model, ambient material is swept up by a radial wide-angle magnetized wind. 
If the wind has an
axial density gradient, the
core of the wind can take on the appearance of a collimated jet
\citep{Shu1995,Ostriker1997,Shang1998}. 
If thermal pressure is unimportant, the
molecular outflow is
momentum-driven  \citep{Shu1991,Masson1992,Li1996b,Matzner1999};
the resulting structure and
kinematics are found to be able to describe e.g.,
HH 111 CO outflow \citep{Nagar1997}.
If there is little or no cooling of the shocked wind and ambient gas,
the dynamics of the outflow can be
different from the momentum-driven shell 
model \citep{Frank1996,Delamarter2000}.

In this paper, 
we present comparisons of
two-dimensional hydrodynamical simulations of both jet- and
wind-driven outflow models with observed molecular outflows in order 
to test which aspects of each model provides a good fit 
to the observed kinematics.
We also present the first quantitative comparison of
the structure and kinematics of the swept-up shells found
in the simulations (which is identified as a molecular outflow)
to new and existing analytical models of jet-driven bow shocks and wide-angle
winds, respectively.
Since we concentrate on the dynamics of the large-scale shell,
our simulations are performed with the ZEUS-2D hydrodynamical code
using approximate cooling rates, and moderate numerical resolution.
We calculate the PV diagrams as well as mass-velocity relationships 
for the shells and compare them
with high-resolution CO observations of a number of molecular outflows.
Calculations are presented for both
steady and time-variable (pulsed) jets and winds.
Our primary goal is to use the simulations to provide a survey of the dynamics
of outflows for a wide variety of models (winds versus jets, pulsed versus
steady), in order to test which models best fit the observations.

\section{Equations and Numerical Method}
The two-dimensional hydrodynamic code, 
ZEUS 2D \cite[developed by][]{Stone1992},
is used to solve the equations of hydrodynamics, 
\begin{equation}
\frac{\partial{\rho}} {\partial{t}}+ \nabla\cdot(\rho{\bf v}) =0
\end{equation}
\begin{equation}
\frac{\partial{\rho {\bf v}}} {\partial{t}}
+\nabla\cdot(\rho {\bf v}{\bf v}) =
-\nabla p + \rho \nabla \Phi
\label{eq:mom}
\end{equation}
\begin{equation}
\frac{\partial{e}} {\partial{t}}
+\nabla\cdot(e{\bf v}) =
-p\nabla\cdot{\bf v}-n^2\Lambda
\end{equation}
where $\rho$, ${\bf v}$, $p$, $e$, and $n$ are the mass density,
velocity, thermal pressure,
internal energy density, and hydrogen nuclei number density,
respectively. 
Some of our simulations use a stratified ambient density $\rho$;
in this case, $\Phi$ is the fixed 
gravitational potential of the ambient material.
We adopt an optically thin
radiative cooling function $\Lambda$ for interstellar gas, with 
the function at high temperature from \citet{MacDonald1981} and 
the function at low temperature from \citet{Dalgarno1972}.
Since we focus on the dynamics,
we assume an equilibrium cooling rate in the simulations, in which the
ionization fraction of hydrogen is calculated by equating the ionization
rate with the recombination rate of hydrogen.
Simulations with nonequilibrium cooling rate have
been performed by e.g., \citet{Stone1993a,Stone1993b}.
Molecular cooling is not included. 
In our simulations, helium is included as a neutral component
with $n(\textrm{He})=0.1\cdot n$, so that $n=\rho/(1.4\cdot m_H$),
where $m_H$ is the mass of atomic hydrogen.
An ideal gas equation of state $P=(\gamma-1)e$ 
with $\gamma=5/3$ is used for the thermal pressure.
Details of the numerical grid and boundary conditions used for the jet and
wind models are given in \S{}\S{} 3.1 and 4.1, respectively.

\section{Jet Model}
\subsection{Numerical Simulation Parameters}
In our standard model, presented herein,
we introduce a perfectly collimated jet into a cold and uniform cloud
with a density of $\rho=1.6\times10^{-20}$ g cm$^{-3}$
(i.e., $n=7\times10^3$ cm$^{-3}$) and a temperature of 30 K.
The jet is circular with a radius of $R_j= 
2.5\times$10$^{15}$ cm \cite[the typical radius of observed optical HH
jets; see][]{Ray1996}, a temperature of 270 K,
and a density of $\rho_j=1.6\times10^{-20}$ g cm$^{-3}$
\cite[i.e., $n=7\times10^3$ cm$^{-3}$, see][]{Gueth1999}.
Consequently,  the ratio of the jet density to ambient density, $\eta$, is 1.
Speeds in observed molecular jets range from 100 to
500 km s$^{-1}$, derived from the proper motion of H$_2$ 1-0
S(1) knots (Micono et al 1998, Coppin et al. 1998). We take the lower
limit and adopt $v_j= 120$ km s$^{-1}$ for the jet speed, resulting in 
a mass loss rate of 1.2$\times10^{-7}$ M$_\odot$ yr$^{-1}$.

The simulations 
are performed in cylindrical coordinates.
We use a computational domain of dimensions 
$(R,z)=(1.25\times10^{16}, 1.25\times10^{17})$ cm and a uniform grid of
$60\times600$ zones, giving a resolution of $2.08\times10^{14}$ cm, or 12 grid
zones per jet radius.
Reflecting boundary conditions are used along the inner $R$
and inner $z$ boundaries, while outflow boundary conditions are used along the
outer $R$ and outer $z$ boundaries. For $R \le R_j$ along the inner $z$
boundary, inflow boundary conditions are used to introduce the jet into the
ambient medium.

We have performed a large number of other simulations in which we vary
$\eta$, $R_j$, and $v_j$, and find results which are qualitatively quite
similar to those presented here.

\subsection{Simulation Results of a Steady Jet}

Figure \ref{fig:jetDTPV} shows the number density, 
temperature and pressure in the simulation at 650 years,
with vectors showing the velocity structure.
The solid line on the density distribution is calculated from a ballistic
bow shock model described in a companion paper \citep{Ostriker2000} 
and is discussed in the next section.
With $\eta=1$, the bow shock speed is about 60 \vkm{}.
Therefore, the jet travels a distance of 1.21$\times10^{17}$ cm 
in 650 years.
Two shocks, a jet shock and a bow shock, 
are clearly seen at the head of the jet in both the
temperature and pressure distributions.
The bow shock interacts with the ambient material 
and forms a thin shell of shocked ambient material
surrounding a high-temperature cocoon (consisting of shocked jet gas)
which in turn
surrounds the jet. The temperature of the shell decreases
from 10,000 K at the jet head to about 100 K near the central
source. 
In this simulation, only simplified cooling is included.
In fact, several 
other processes, such as molecular cooling,
collisional ionization of hydrogen,
recombination and dissociation of molecular hydrogen, and
formation of molecular hydrogen on the surfaces of dust grains,  
would also affect the gas temperature determination in the shell. 
Simulations
including these processes have been presented by 
\citet{Suttner1997}.
In their simulations, there is a significant fraction of
molecular hydrogen in the shell. Therefore,
if molecular cooling were included in our simulation, 
the temperature of the shell would actually be lower.
Regions of the shell with temperature lower than a few hundred K can be
identified with the observed CO molecular outflow, whereas regions of
the shell with temperature above 1,000 K
will produce the observed H$_2$ bow shock.

Hydrodynamical simulations have been presented by many authors
\cite[e.g.][]{Blondin1989,Blondin1990,Gouveia1993,Stone1993a,Stone1993b,
Stone1994,Biro1994,Suttner1997,Smith1997,Downes1999}
using different cooling functions and different resolutions.
Some of the simulations are at higher resolution than this work and thus
show much more small scale structure.
However, the overall shape and size of the shell structure in higher
resolution simulations presented elsewhere, and also computed by ourselves,
is similar to the results we present here, 
indicating that the underlying kinematics is not much different.
We will further discuss the effects on the shell 
due to different numerical resolution later in this section.
The goal of this paper is to study the overall large-scale kinematics 
of the shell in comparison to CO observations. Detailed high-resolution
studies of propagating jets necessary for understanding small-scale features
in H$_2$ and optical observatoins are presented in the papers referred to 
above.

The velocity of the material in the shell is almost perpendicular
to the shell surface. However, the
velocity of the material in the cocoon is mostly parallel to the jet axis,
indicating that the material in the cocoon mostly comes from the jet. 
The shell structure is corrugated because of the variations in the diameter
of the jet as it propagates, induced by interaction with shocked material in
the cocoon \cite[e.g.][]{Blondin1990}.
At higher resolution, the shell is more corrugated but the overall structure
is still the same (see the discussion in next paragraph).

Figure \ref{fig:jetDTP} 
shows the number density, temperature and pressure along an axial 
cut through the head of the jet.
The jet shock and bow shock are clearly seen in the temperature and pressure
distributions.
The shocked ambient material and jet material are compressed by the shocks
so that the number density peaks between the two shocks.
The maximum temperature is 10$^4$ K at the shock fronts.
The immediate postshock temperature
is $T_s = 3 \bar{m}v_s^2/16k$,  where $\bar{m}$ is
the average mass per particle, or about 10$^5$ K for $\gamma=5/3$.
For a strong shock, the postshock speed is about one-fourth of the shock
speed (for $\gamma=5/3$). The immediate postshock
cooling length can be approximated as \cite[see][]{Blondin1990}, 
\begin{equation}
l_{cool} \approx \frac{v_s t_{cool}}{4}=\frac{v_s e}{4 n_s^2 \Lambda(T_s)}=
\frac{3 v_s k T_s}{8 n_s \Lambda(T_s)}
\end{equation}
where $v_s$ and $n_s$ are the shock velocity and postshock number density. 
Since the density increases rapidly toward the interface (contact
discontinuity) of the two shocked layers, 
the cooling rate actually increases rapidly
toward the interface.
For a postshock number density from 10$^4$ to 10$^6$ \cm3{},
the cooling length due to ionic cooling
varies between 10$^{13}$ to 10$^{11}$ cm. 
This length is not resolved in our
simulation, therefore the postshock temperature drops to 10$^4$ K 
immediately behind the shocks. 
The cooling length behind the shocks in the simulation
is thus constrained by the numerical resolution, giving
a length of about 10$^{15}$ cm.
Consequently, the shock thickness and thus the shell thickness
decrease with increasing resolution.
In order to investigate the effect of numerical resolution of the dynamics
of the swept-up shell which surrounds the jet cocoon, we have performed a
series of simulations with resolution from $4\times10^{14}$ to
$5\times10^{13}$ cm. We find the shape and kinematics of the shell is
controlled primarily by the flux of transverse momentum associated with dense
gas ejected radially from the shocked jet and ambient gas at the head of the
jet. We find the density of the material ejected from the jet head
increases with increasing resolution, so that the transverse momentum flux
from the jet head does not change significantly with resolution.
Thus, even though the shell thickness
decreases with increasing resolution, 
the shell structure and kinematics
do not depend on the resolution. 
Therefore, our simulation can be used for studying the
shell structure and kinematics of the jet-driven outflows.



\subsection{Comparison between the Simulation and a Ballistic Bow Shock Model}
In this section, we compare our simulation with the ballistic 
bow shock model developed in a companion paper \citep{Ostriker2000}. 
In the model,
the shell structure and kinematics are determined by the transverse momentum
flux from the working surface surrounding the hot shocked material
at the jet head.
The transverse momentum flux can be expressed as the ambient mass flux
flowing into the working surface
times a factor proportional to the sound speed $c_s$ behind  shocks, 
i.e., $\dot{P}_{oR}= \beta \pi R_j^2 v_s \rho c_s$, 
where $\beta$ is a constant and $\rho$ is the
undisturbed ambient density \citep{Ostriker2000}. 
In the simulation, the
temperature drops to 10$^4$ K immediately behind the shocks, giving 
a sound speed of 8 \vkm{}. Direct measurement of the transverse
momentum flux at $R=R_j$ in the simulation gives $\beta=2.2$. The value of
$\beta$ increases with distance from the working surface due to the
finite pressure gradient in the shell; also
the effective radius of the working surface is slightly 
greater than the jet radius.
A larger value of $\beta$ is therefore needed to fit the simulation.
As can be seen,
the shell structure is reasonably described by
the model with $\beta =4.1$.
As a result,
the shell in our simulation can be expressed as $z\propto R^{3}$ 
\citep{Ostriker2000}. This shell structure is consistent 
with that found in the simulations of \citet{Smith1997} and \citet{Downes1999}.

Figure \ref{fig:jetP} presents a comparison of the
velocity structure in the simulation
and the ballistic bow shock model.
Figures \ref{fig:jetP}\tlabel{a} and \ref{fig:jetP}\tlabel{b} show
the velocity fields in a frame of reference moving with the bow shock
for the simulation and model, respectively.
In both panels, the velocity is plotted as arrows over a gray-scale image of
the pressure distribution taken from the simulation.
In the simulation (Fig. \ref{fig:jetP}\tlabel{a}),
the hot shocked material at the jet head flows transversely into the shell. 
As the material moves along the shell, part of the material 
flows into the cocoon from the inner surface of the shell
because of the effect of thermal pressure within the shell.
The velocity along the
inner surface of the shell is smaller than the outer surface, producing 
a velocity gradient across the shell.
This gradient is a consequence of the fact that the 
ambient material being swept-up by the shell has zero transverse momentum.
The bow shock provides an initial transverse impulse; however, the 
swept-up ambient material does not immediately mix with
high velocity shocked jet gas ejected from the head of the jet,
therefore producing the velocity gradient
across the shell as seen in Figure \ref{fig:jetP}\tlabel{a}. 
In Figure \ref{fig:jetP}\tlabel{b}, we plot
the velocity of the \nam{} material (open arrows) 
and the mean velocity of the material in the shell (closed arrows) 
from the bow shock model \cite[see][]{Ostriker2000}. The newly swept-up
material lies along the outer surface of the bow shock. 
The velocity of the \nam{} material
has larger magnitude than the mean velocity of
the material in the shell, indicating that
mixing in the shell is not
immediate, giving rise to the velocity gradient in the shell as seen
in Figure \ref{fig:jetP}\tlabel{a}.
Figures \ref{fig:jetP}\tlabel{c} and \ref{fig:jetP}\tlabel{d} show the
velocity fields for the simulation and model, respectively, 
in the observer's frame plotted as arrows over a gray-scale image of the 
density distribution from the simulation.
In the simulation (Fig. \ref{fig:jetP}\tlabel{c}), it is clear that
there is a gradient in the velocity across the shell between the inner and
outer shell surfaces,
with the outer shell having smaller and less forward velocity than 
the inner shell. In the model (Fig. \ref{fig:jetP}\tlabel{d}), the velocity
of the \nam{} material (open arrows) 
is also smaller and less forward than the mean velocity of  
the material in the shell (closed arrows), consistent with the simulation.

Figure \ref{fig:jetvrz} shows transverse velocity, $v_R$, and longitudinal
velocity, $v_z$, of the material in the shell in the observer's frame for
both the simulation (images) and bow shock model (lines).
The solid lines are calculated with the mean velocity of the shell material
and the dashed lines are calculated with the velocity of the \nam{} 
material from the bow shock model.
Both $v_R$ and $v_z$ drop quickly away from
the tip of the bow shock, because momentum in the shell is shared 
with the \nam{} ambient material.
The maximum transverse velocity is about 10 \vkm{}, comparable to the 8
\vkm{} sound speed at 10$^4$ K as
predicted by the ballistic bow shock model.
The transverse velocity of the
outer-shell swept-up ambient material is nearly the same 
as the mean transverse velocity of the material already in the shell.
Therefore, except near the tip,
the dashed and solid lines both match
the transverse velocity of the simulation reasonably well. 
As for the longitudinal velocity,
the solid line (representing mean $v_z$)
matches the high velocity better than the dashed line, while
the dashed line (representing \nam{} material)
matches the low velocity better than the solid line.
This is because mixing is more complete near the jet head,
so that the high velocity is similar to the mean velocity. However,
there is almost no mixing in the wing.
In addition, the material from the working surface partly flows into the 
cocoon, further reducing the mixing. The velocity of the shell
material in the wing is thus similar to that expected for \nam{} material.

Overall, the simple ballistic bow shock model is a good fit for the shell
shape and the transverse velocity of the shell material.
Since the material from the working surface partly flows into the cocoon 
as it moves up the shell and the mixing of the \nam{}
material with material already in the shell is not complete, the model
only provides lower and upper limits for
the longitudinal velocity of the
shell material. The lower limit is set by the velocity of the \nam{}
material, whereas the upper limit is set by the mean velocity of the shell 
material.
If the numerical resolution were much higher so that the shear layer between
the swept-up and shell material were resolved, mixing might be increased.
Moreover, if molecular cooling were included to lower the thermal pressure
in the shell, there would be less material 
flowing into the cocoon. 
In that case, the shell material would have a higher forward velocity.
Previous simulations which include molecular cooling
\cite[e.g., ][]{Smith1997,Downes1999,Volker1999}
indeed show the shell material that has higher forward velocity than that in our
simulation. 

\subsection{PV Diagrams and MV Relationship} 
In our simulation, only a simplified treatment of radiation
cooling is included, and moreover, the
cooling length in the shell is often not resolved. Therefore,
the temperature of the shell material can not be used to calculate line
emission. Instead, our kinematic diagnostics
are based on the mass rather than the line emission.
Since we identify the shell in the simulation
as the molecular outflow \citep{Lee2000},
we only focus on the shell kinematics and
thus exclude the jet and cocoon material from our calculations.
Since the density contrast between the shell and cocoon material
is large (see Figure \ref{fig:jetDTPV}), we can define
a boundary between the shell and cocoon material and
mask out the cocoon and jet material.
Unavoidably, a little cocoon material close to the shell will be
included in our calculations; however, the mass contribution of that cocoon
material is small.

Figure \ref{fig:jetPV} shows position-velocity (PV) diagrams for the shell
in the simulation cut along the outflow axis
at three inclinations ($i$) to the plane of the sky. 
The jet and cocoon material are masked and
excluded from the calculations.
The PV diagrams are calculated by projecting the 2-dimensional density
distribution into 3-dimensions and summing the mass
along each line of sight for any given velocity.
The solid lines are PV diagrams calculated from the ballistic 
bow shock model using the mean velocity of the shell material, while
the dashed lines using the velocity of the \nam{} ambient material.
It is clear that
the shell material in the wings is dominated by the unmixed shocked ambient
material, so that the observed velocity is similar to the
observed velocity of the \nam{} ambient material. 
The PV diagrams are similar to that 
seen in other simulations: they consist of convex spur structures 
with the highest velocity at the tip \citep{Smith1997,Downes1999}.
At $i=0$\degree, where the  outflow axis is in the plane of the sky, 
the diagram is symmetric about zero velocity. 
As the inclination increases (with the jet pointing away from the observer), 
the blueshifted shell material and low-velocity redshifted shell
material moves toward zero velocity, 
while the high-velocity redshifted shell material moves
away from zero velocity. Therefore
at $i=60$\degree, the blueshifted shell material and low-velocity
component of redshifted shell material merge with the ambient
material, leaving the high-velocity 
shell material mostly visible on the redshifted side.

Figure \ref{fig:jetmv} shows mass-velocity (MV) relationship for the shell
material at the same three inclinations. The MV relationship is computed by
summing the mass for any given velocity.
Both the redshifted (open square) and blueshifted (filled square)
masses are shown.
The curves in the figure are not smooth since the shell itself is not smooth.
The dashed lines are the fits to the redshifted mass with a power-law mass
velocity relationship, $dM/dv_{obs}\equiv m(v_{obs}) \propto v_{obs}^{-\gamma}$, where the
power-law index, $\gamma$, is indicated at the upper right corner in each panel. 
As found by \cite{Downes1999}, our result shows that $\gamma$ depends strongly on the
inclination;  $\gamma$ is 3.44 at $i=0$\degree{} and
decreases to 2.01 at $i=60$\degree. 
$\gamma$ is also consistent with that found by \citet{Smith1997}
if their channel line
intensity scales with $dm(v)$ rather than $v^2 dm(v)$.
Notice that what is actually observed is a
variation in CO line intensity with velocity.
If the CO line  is optically thin
and the temperature of the gas is constant and
higher than the excitation temperature of the line,
CO line intensity is directly proportional to mass 
\citep[see, e.g.][]{McKee1982}.
Therefore, our mass-velocity relationship can only be applied to the 
observations if the CO line is optically thin and the temperature of the
shell is constant. Readers should consult e.g. \citet{Smith1997} for
a more sophisticated calculation of CO line intensity-velocity relationship.

Since the bow shock model \citep{Ostriker2000}
can reasonably reproduce the transverse velocity in
the simulation, the model can also account for 
the MV relationship at $i=0$, where the observed velocity depends on the
transverse velocity only.
The solid line at $i=0$\degree{} in Figure \ref{fig:jetmv} 
is calculated from the bow shock model and matches
the MV relationship of the simulation well.
\subsection{Comparison with a Pulsed Jet Simulation}
Optical and infrared observations reveal that protostellar 
jets  usually consist of a series of well aligned knots and internal 
bow shocks moving away from the source at high velocity 
\citep{Coppin1998,Micono1998}. 
These knots and bow shocks are thought to be associated with
internal working surfaces produced by time variability in the jet
\citep{Reipurth1988,Raga1990}. In this section, 
we present a simulation of a time-variable (pulsed) jet
and compare it with the steady jet simulation. 
The jet velocity and density are assumed to be
\begin{equation}
v_j' = v_j(1+A \sin \frac{2\pi t}{P})
\end{equation}
\begin{equation}
\rho_j' = \rho_j/(1+A \sin \frac{2\pi t}{P})
\end{equation}
where $A$ is the amplitude of the variation and $P$ is the period. The
momentum of the jet is constant over the variation. 
We use $A=0.5$ and $P=310$ years, allowing
a strong variation of the jet momentum to 
demonstrate the effects of the internal working surface.

Figure \ref{fig:pjetden} shows a gray-scale image of the density distribution
in the simulation at 610 years, along with
PV diagrams and the MV relation at two inclinations.
At a time of 310 years, two internal working surfaces are formed in the
jet, producing a knot and an internal bow shock structure. 
Due to thermal pressure gradients, shocked jet material is ejected 
radially from each internal working surface into the cocoon.
The ejected material initially appears as a knot. 
As the working surface travels down the jet axis, 
the knot grows into an internal bow shock surface and eventually 
interacts with the shell. 
Since the inertia of the shell is large, 
the shell structure is not significantly affected by the internal 
bow shock \cite[see e.g.,][]{Stone1993b,Biro1994,Suttner1997}.
The internal bow shock, which consists of jet material, has 
a higher forward velocity than the leading bow shock surface, because it
interacts only with the low-density, forward-moving cocoon and not the 
dense, stationary ambient gas.
From the PV diagram at $i=0$\degree{}, it is clear that the velocity of the
internal bow shock surface decreases slowly with $R$ before it 
interacts with the shell, again
because there is not much material in the cocoon.
As the bow shock surface interacts with the shell, 
momentum is transfered to the shell material and the velocity of the shell
increases. As a result, the velocity structure of the shell 
is broken at the interaction point into two convex spur PV structures, 
one associated with the leading bow shock and one 
with the internal bow shock.
The internal bow shock surface has a higher forward 
velocity and appears as high velocity gas in the PV diagram at
$i=30$\degree. The mass at high velocity in this simulation
is higher than that in the steady jet simulation. 
Therefore, the mass-velocity relationship in the pulsed jet simulation
has a smaller power-law index than that in
the steady jet simulation (compare Figures \ref{fig:jetmv} and \ref{fig:pjetden}).

\subsection{Steady Jet in a Highly Stratified Ambient Medium}
We have also computed a steady jet in a highly stratified ambient medium 
with density distribution (in cylindrical coordinates)
\begin{equation}
\rho = \frac{\rho_c}{1+(z/z_c)^2}
\end{equation}
where $\rho_c$ is the density at $z=0$
and $z_c$ is the effective core-radius of the ambient medium.
We use $\rho_c=1.6\times10^{-19}$ g cm$^{-3}$
(i.e., $n=7\times10^4$ cm$^{-3}$) and $z_c=$1.25$\times$10$^{17}$ cm. 
The jet is underdense ($\eta=0.1$) at $z=0$ with $\eta$ increasing as the jet
propagates down the jet axis.
Due to the strong cooling of the shocked material at the head of
the jet and the cocoon material, we find
the structure and kinematics of the swept-up shell in an
underdense jet are more or less ballistic and
similar to that of the steady jet in an uniform ambient material with
$\eta=1$ (see \S 3.2). Thus, stratification of the ambient medium above does
not modify the kinematic signature of the jet-driven outflow.

\section{Wide-angle Wind Model}
\subsection{Numerical Simulation Parameters}
The wide-angle wind model is an alternative to the bow shock of a collimated
jet. In this model,  a molecular outflow is ambient
material swept-up by a radial wind from a young star.
The wind could be stratified in density if it
emerges from an extended
disk \cite[e.g.][]{Ostriker1997} or by the action of latitudinal magnetic
stresses in a rotating protostellar x-wind \citep{Shu1995}.
In a spherical coordinate system ($r,\theta,\phi$), \citet{Shu1995} 
have shown that for an
x-wind model the wind density can be approximated as
\begin{equation}
\rho_w\propto 1/(r\sin\theta)^2\;.
\end{equation}
This is also true for more general force-free winds 
\citep{Ostriker1998,Matzner1999}.
We have performed a number of time-dependent 
hydrodynamical simulations of the evolution of 
wide-angle wind in stratified ambient medium. The wind is introduced as a
boundary condition along a spherical shell of radius $r_w$, with density
distribution
\begin{equation}
\rho_w = \frac{\rho_{wo}}{r_w^2(\sin^2\theta+\epsilon)}
\end{equation} 
where $\rho_{wo}$ is a constant and $\epsilon$ 
is a small value avoiding a singularity of the wind density
at the pole ($\theta=0$).
In the x-wind model, the wind velocity $v_w$ is approximately the same at
all angles \citep{Najita1994}. 
Here, we assumed the wind velocity decreases toward the equator 
($\theta=\pi/2$) and is given by
\begin{equation}
v_w = v_{wo} \cos\theta
\end{equation}
where $v_{wo}$ is the velocity at the pole.
The mass loss rate of the wind is set to that in the jet
simulations, thus
\begin{equation}
4 \pi \int_0^{\pi/2} \rho_w v_w r_w^2 \sin \theta d\theta = 2 \rho_j \pi R_j^2 v_j
        = 1.2\times 10^{-7} M_\odot\; yr^{-1}
\end{equation}
Letting $v_{wo}=v_j$, $r_w=R_j$ and $\epsilon = 0.01$, we have
$\rho_{wo}=3.5\times10^{-21} r_w^2$ g cm$^{-1}$, giving
$\rho_w=3.5\times10^{-19}$ g cm$^{-3}$ (i.e., $n=1.5\times10^5$ 
cm$^{-3}$) at the pole, and $\rho_w=3.5\times10^{-21}$ g cm$^{-3}$ 
(i.e., $n=1.5\times10^3$ cm$^{-3}$) at the equator.

For the ambient medium, a flattened torus with density 
\begin{equation}
\rho_a = \frac{\rho_{ao} \sin^2 \theta}{r^2}
\end{equation}
is used, where $\rho_{ao}$ is a constant.
This density distribution is appropriate for magnetized
cores \cite[see e.g.,][]{Li1996b}. In the simulations,
$\rho_{ao}$ is set to $1.6\times10^{-18} r_w^2$ g cm$^{-1}$, which leads
$\rho_a = 1.6\times10^{-18}$ g cm$^{-3}$ (i.e., $n=6.8\times10^5$
cm$^{-3}$) and $\eta=2.2\times10^{-3}$ at the equator when $r=r_w$.
We also added
a fixed gravitational potential $\Phi$
to keep the ambient material stable.
This is the most favorable
distribution of the ambient material for the wind model.
Analytically, it has been shown to produce
not only the correct mass-velocity relationship, 
but also the collimation required for molecular outflows
\citep{Li1996b,Matzner1999}.

In the x-wind model,
the wind and ambient material are actually magnetized
\citep{Shu1994,Shu1995,Li1996b}. For a magnetized gas with a low ionization fraction, 
a non-dissociative C-shock can occur for shock velocities
up to 40 $-$ 50 \vkm{} \citep{Hollenbach1997}. Therefore, a C-shock
interaction is likely to take place in the wings of the outflow 
lobe, with a J-shock interaction near the tip
of the lobe. At temperatures below 10$^4$ K,
the cooling of the shocked material in the wings
is thus dominated by molecular cooling
\citep{Hollenbach1997}.
Since molecular cooling is very efficient for the density used in our
simulation, an isothermal state function is used for the thermal pressure.
The effect of magnetic fields on the dynamics of wide-angle winds
will be presented in a future publication. 

The simulations of this model are performed in a spherical coordinate
system. 
We use a computational domain of dimensions 
$(r,\theta)=(2.5\times10^{15} - 1.25\times10^{17} \textrm{cm}, \pi/2)$ and a uniform grid of
$600\times400$ zones, giving a resolution of $2.04\times10^{14}$ cm in $r$ 
and 4$\times10^{-3}$ radians in $\theta$.
Reflecting boundary conditions are used along the inner $\theta$
(where $\theta=0$)
and outer $\theta$ (where $\theta=\pi/2$) boundaries, while outflow boundary conditions are used along the
outer $r$ boundary. Inflow boundary conditions are used for inner $r$ boundary 
to introduce the wind into the ambient medium.

\subsection{Simulation Results for an Isothermal Steady Wind}

Figure \ref{fig:windiden} presents the density distribution of an isothermal
wind on a logarithmic scale at 390 years, with vectors 
showing the velocity structure in the shell. The
solid line indicates the predicted shape
from a simple momentum-drive shell model derived in the next section.
In the simulation, the temperature is set to 100 K, a
typical temperature of the CO outflow in the observations \citep{Fukui1993}.
The central high density ``jet'' spreads with a small opening angle
because we have used a small factor, $\epsilon$,
to avoid the singularity of the wind density at the pole.
The shell is elongated in the polar direction, 
due to the stratification of the ambient material 
and the variation in angle of the wind thrust. The shell everywhere expands 
in the radial direction.
At a temperature of 100 K, the thermal pressure in the shell
is negligible compared to the ram pressure of the wind. 
Consequently, the shocked wind and ambient materials
merge into a thin shell of shocked material.

\subsection{Comparison Between the Simulation and A Momentum-driven Shell Model}
Assuming immediate cooling behind shocks and no relative motion 
of the shocked material, \citet{Shu1991} and \citet{Li1996b}
derived the shape and 
kinematics of the swept-up shell by considering mass and momentum 
conservation in each angular sector.
In their calculations, the wind mass added to the shell is not included.
Following the same approach, we rederive the shape and kinematics 
of the shell including the wind mass added to the shell below. 
The mass (ambient mass $+$ wind mass) and momentum fluxes added 
to the shell per second per steradian 
are
\begin{equation}
\frac{dM_s}{dt} = r^2 \rho_a v_s + r_w^2 \rho_w(v_w-v_s)
\label{eq:massw}
\end{equation}
\begin{equation}
\frac{d (M_s v_s)}{dt} = r_w^2 \rho_w(v_w-v_s)v_w
\end{equation}
where $M_s$ is the mass flow into the shell and $v_s$ is the shell velocity.
Since the ambient material has a density varying as $r^{-2}$,
the dilution of the wind ram pressure is exactly compensated
as the outflow expands away from the source, 
resulting in a swept-up shell which proceeds outward with a
constant velocity along any radial line \citep{Shu1991,Masson1992}.
Therefore $v_s$ is constant and
\begin{equation}
v_s\frac{dM_s}{dt} = r_w^2 \rho_w(v_w-v_s)v_w
\label{eq:momw}
\end{equation}
Substituting equation (\ref{eq:massw}) into equation (\ref{eq:momw})
and solving for $v_s$, we have
\begin{equation}
v_s = \frac{v_w}{1+\eta_w^{-1/2}}
\end{equation}
where $\eta_w\equiv\frac{r_w^2 \rho_w}{r^2 \rho_a}$. 
The shell structure is then given by
\begin{equation}
r_s = v_s t + r_w = \frac{v_w t }{1+\eta_w^{-1/2}} + r_w
\end{equation}
The maximum width of the shell can also be found at angle
$\sin\theta=[2+(\frac{\rho_{ao}}{\rho_{wo}})^{1/2}]^{-1/2}$ as
\begin{equation}
W_w \approx v_{wo} t (\frac{\rho_{wo}}{\rho_{ao}})^{1/4} 
\end{equation}
for $W_w \gg r_w$ and $\epsilon=0$ . 
Since the length of the shell is $\approx v_{wo} t$,
the width-to-length ratio is about
$(\frac{\rho_{wo}}{\rho_{ao}})^{1/4}$ and constant with time.
The ratio of the total longitudinal momentum flux to the total 
transverse momentum flux in the shell can be expressed as
\begin{equation}
\mathcal{R} \approx \frac{{-1+\frac{1+a}{\sqrt{a}}} \arctan \frac{1}{\sqrt{a}}}
    {{-1+\frac{\sqrt{1+a}}{2}} \log \frac{\sqrt{1+a}+1}{\sqrt{1+a}-1}}
\end{equation}
where $a=\sqrt{\frac{\rho_{wo}}{\rho_{ao}}}$. 
In our simulation, $(\frac{\rho_{wo}}{\rho_{ao}})^{1/4} \approx 0.22$ and
$\mathcal{R} \approx 4.5$,  therefore,
the wind-driven shell is elongated in the polar direction.

The solid line in Figure \ref{fig:windiden} indicates the
shell shape calculated from the momentum-driven shell model. 
As can be seen, the model provides a good fit.
Due to finite thermal pressure gradients,
the shell in the simulation is a little larger than the model.
Figure \ref{fig:windivrz} shows the transverse velocity, $v_R$, and the
longitudinal velocity, $v_z$, along the shell, with solid lines indicating
the predicted velocities from the model. Again, the model provides a good
description of the velocity structure. The transverse velocity increases
from zero at the source and then decreases to zero at the tip,
while the longitudinal velocity increases linearly with distance.
In contrast to the jet simulation, the transverse velocity
vanishes at the
tip because the thermal pressure is negligible 
in the isothermal simulation at such a low temperature. 

Comparing Figure \ref{fig:windiden} with Figure
\ref{fig:jetDTPV},
we see that the wind-driven shell is wider than the jet-driven shell,
although the wind and jet both have the same mass loss rate
in the simulations. The width-to-length ratio for the jet-driven shell
is $2(3\beta c_s/v_s)^{1/3}|R_j/z|^{2/3}$ and decreases with 
increasing $z$ \citep{Ostriker2000}. 
The width-to-length ratio for the
wind-driven shell, on the other hand,
is $(\frac{\rho_{wo}}{\rho_{ao}})^{1/4}$ and 
independent of $z$. Therefore, even with optimized ambient medium
stratification most favorable for {\it narrow} outflows in the wind model 
and {\it wide} outflows in the jet model, the wind-driven shell always becomes
wider than the jet-driven shell beyond a certain distance of $z$;
in our case, this occurs at about 30 $R_j$.

\subsection{PV Diagrams and MV Relationship}
Figure \ref{fig:windipv} shows PV diagrams for 
the shell material cut along the outflow axis
at three inclinations, with lines indicating the 
model calculations. 
The PV diagrams are significantly different from that seen in
the jet simulations,  
showing a lobe structure that is tilted with inclination
such that primarily red or blue, but not {\em both},
emission is seen if the inclination is nonzero.
At an inclination $i$,
the projected distance of the shell material is 
\begin{equation}
d = r_s (\cos \theta \cos i - \sin\theta\sin i \cos \phi)
\end{equation} 
and the observed velocity is
\begin{equation}
v_{obs} = v_s (\sin \theta \cos \phi \cos i + \cos \theta\sin i)
\end{equation}
where $\phi$ is the azimuthal angle in the shell. Therefore the PV diagram
along the major axis is given by
\begin{equation}
d = (\frac{v_{obs} t}{\sin \theta \cos \phi \cos i + \cos \theta\sin i}+r_w)
    (\cos \theta \cos i - \sin\theta\sin i \cos \phi)
\end{equation} 
where $\phi=0$ for the front wall and $\phi=\pi$ for the back wall of the
shell.
As can be seen in Figure \ref{fig:windipv}, 
the PV diagrams from the simulation are well fit by the model.

Figure \ref{fig:windimv} shows the mass-velocity (MV) relationship 
for the shell material. Both the redshifted (open square) and 
blueshifted (filled square) masses are shown. 
The dashed lines are the fits to the redshifted mass with a power-law mass
velocity relationship, $m(v_{obs}) \propto v_{obs}^{-\gamma}$, where the
power-law index, $\gamma$, is indicated at the upper right corner in each panel.
The solid lines are from the model calculations.
\noindent
The swept-up mass (ambient mass $+$ wind mass) per steradian, $m(\mu,\phi)$, is
\begin{eqnarray}
m(\mu,\phi) &=& \int \frac{dM_s}{dt} dt
       = r_w^2 \rho_w v_w \eta_w^{-1/2}  t
\end{eqnarray}
where $\mu= \cos\theta$.
Therefore, the mass per unit velocity is
\begin{eqnarray}
m(v_{obs}) &=& \int m(\mu,\phi) \frac{d\phi}{dv_{obs}}  d\mu \nonumber \\
       &=& \int m(\mu,\phi)\frac{- d\mu}{v_s \sin \theta \sin \phi \cos i}  \nonumber \\
       &=& \int \frac{- r_w^2 \rho_w v_w \eta_w^{-1/2} t d\mu}
             {\sqrt{v_s^2 \sin^2 \theta \cos^2 i - (v_{obs}-v_s \cos \theta \sin i)^2}}  
\end{eqnarray}
In the simulation, the wind sweeps up a slightly larger shell than predicted
by the model (see Figure \ref{fig:windiden}), therefore,
the mass from the model is multiplied 
by a factor of 1.3 to match the simulation.
Again, the model matches the MV relationships very well.
The slope of the curve is steeper at higher velocity.
Fitting to the mass beyond a few \vkm{} 
gives a $\gamma$ from 1.3 to 1.8, a little smaller than
the original x-wind model, which is about 2 
\cite[see, e.g.,][]{Li1996b,Matzner1999}.

\subsection{Comparison with an Isothermal Pulsed Wind Simulation}

We have also computed the evolution of an isothermal time-variable 
(pulsed) wind for comparison to the simulation of a isothermal steady
wind. As in the pulsed jet simulation, the wind velocity and  
density are assumed to be
\begin{equation}
v_w' = v_w(1+A \sin \frac{2\pi t}{P})
\end{equation}
\begin{equation}
\rho_w' = \rho_w/(1+A \sin \frac{2\pi t}{P})
\end{equation}
where $A$ is the amplitude of the variation and $P$ is the period. Again,
the momentum of the wind is constant over the variation. 
In the simulation, $A=0.5$ and $P=115$ years.

Figure \ref{fig:pwindden} shows a gray-scale image of the density distribution
in the simulation at 296 years, along with 
PV diagrams and the MV relation at two inclinations.
As the wind velocity changes,
the fast wind overtakes the slow wind, producing internal shocks
inside the shell. As was seen in the pulsed jet simulation,
the shell also shields the internal surfaces from directly 
interacting with the ambient material.
The polar region of the internal shocks
moves down the axis without interacting with the shell until it reaches the
tip of the shell. The structure of the internal shocks
is thus determined by the intrinsic angular distribution
of the wind velocity, as indicated by the solid lines, and
is flatter further out from the source. 
This structure of the internal shocks is much flatter than the 
observed internal H$_2$ bow shocks in e.g., HH 212
\citep{Zinnecker1998} and HH 111 \citep{Coppin1998}.
To account for the observed shape of internal H$_2$ bow
shocks in these outflows, the temporal variation in the wind model
would need to be confined to a small angle along the outflow axis, 
and the wind speed would also need to decrease more rapidly toward the
equator.
In the simulations,
the wings of the internal surfaces interact with the shell and transfer
momentum to the shell, producing multiple structures in the PV diagrams, 
each associated with an internal surface.
In the PV diagrams,
the polar region of the internal surfaces appear almost as straight lines.
The power-law index of the MV relation is smaller
than the steady wind simulation, indicating there is more mass at higher
velocity.

\section{Comparison with Observations}

Currently, the best example for comparison with 
the features evident in the
jet-driven model is
probably the HH 212 outflow \citep{Zinnecker1998}. 
In Figure \ref{fig:jetHH212}, the left panel shows CO emission contours
\citep{Lee2000} plotted over an
H$_2$ image \citep{Zinnecker1998}, and
the right panel shows the PV diagram of the CO emission cut along the jet axis
\citep{Lee2000}.
The jet itself is seen as a
series of knots and bow shock structures in H$_2$ emission along the outflow
axis. The CO emission surrounds the H$_2$ emission.
This morphological relationship can be produced in 
a jet simulation, in which the H$_2$ emission is produced near the bow
shock surface and CO emission is produced in the wings and 
around the bow shock surface.
The PV diagram of the CO emission in HH 212
along the jet axis shows a series of convex 
spur structures on both redshifted and blueshifted sides
with the highest velocity near the H$_2$ bow tips,
qualitatively similar to the PV diagram 
at zero inclination in our pulsed jet simulation (see Fig. \ref{fig:pjetden}).
Notice that the PV diagram of the outflow is a 
little asymmetric due to the inclination effect.
At $i=0$\degree{}, the PV diagram in the ballistic bow shock model 
can be expressed as
$v_{obs} = [\beta c_s v_s^2 R_j^2/9]^{1/3} |z|^{-2/3} $, 
where $|z|$ is the projected distance from the bow shock.  
For an internal surface, this relationship becomes
$v_{obs}=[\beta c_s \triangle v (v_{si}-v_e) R_j^2/(18 \rho_e/\rho_j)]^{1/3}|z|^{-2/3}$,
where $v_{si}$ is the internal shock speed, $\triangle v$ is velocity jump 
across the internal shock, $v_e$ and $\rho_e$ are the velocity and density of
the cocoon material \citep{Ostriker2000}.
The solid lines in the Figure \ref{fig:jetHH212} are calculated with 
$[\beta c_s \triangle v (v_{si}-v_e) R_j^2/(18 \rho_e/\rho_j)] = 3650$
arcsec$^2$ (km/s)$^3$ and
$R_j$ in unit of arcsec.
With $R_j=0.2"$ (i.e., 100 AU at 460 pc) and
$\beta c_s=32$ km/s, we have 
$\frac{\triangle v (v_{si}-v_e)}{\rho_e/\rho_j}=5.1\times 10^4$ (km/s)$^2$.
For HH 212, $v_{si}$ is about 75 \vkm{} \citep{Davis2000}.
If $v_{si}-v_e \approx 50$ \vkm{}, 
$\triangle v \approx 50$ \vkm, then $\rho_e/\rho_j =0.05$.  
The close agreement between the observed properties of the HH 212 jet, and
the parameter values needed to fit the observations with the jet-driven
outflow model, leads us to
conclude this molecular outflow may be driven by the underlying jet.

Two other good examples for comparison with the jet-driven model
are the HH 240/241 outflow \cite[][]{Lee2000} and the
NGC 2264G outflow \cite[][]{Lada1996}.
The CO emission of these outflows show bow shock structures.
The PV diagrams of the HH 240/241 (the western lobe)
and NGC 2264G CO outflows along the major axis also show the similar 
PV structures to that seen in our simulations at an
inclination of 60\degree. 
The actual inclinations of these outflows depend on the mixing of 
the shell material.
Smaller inclination is required to reproduce the PV
structures of these outflows if the mixing is more complete.

A good example of the wind-driven model is
the VLA 05487 outflow \citep{Lee2000}. 
In Figure \ref{fig:jetHH110}, the left panel shows the CO emission contours
\citep{Lee2000} plotted over the
H$_2$ image \citep{Garnavich1997}, and the
right panel shows the PV diagram of the CO emission cut along the jet axis
\citep{Lee2000}. Since there are other two outflows to the
south contaminating this outflow, 
we concentrate on the northern lobe.
The northern lobe of the CO outflow forms a conical structure
around jet-like H$_2$ emission. The PV diagram along the major axis
shows a parabolic PV structure extending out from the source. 
The CO emission structure and PV structure have been reasonably
modeled with a radially expanding
parabolic shell with a Hubble-law velocity structure, which is
derived from the x-wind model \citep{Lee2000}.
This outflow is large, extending at least 8' to the north
\citep{Reipurth1991}, the observations
in \citet{Lee2000} only revealed the inner part of the CO molecular outflow.  
Therefore the emission structure and PV structure of this outflow
both appear parabolic. The solid line in the PV diagram is calculated from
our modified x-wind model with $i=18$\degree{}, $t=8100$ years and 
$\frac{\rho_{wo}}{\rho_{ao}}=\frac{1.27\;\textrm{(km/s)}^2}{v_{wo}^2}$
at a distance of 460 pc. This gives $v_o = 0.27$ \vkm{} arcsec$^{-1}$ 
and $C=0.24$ arcsec$^{-1}$ using Equation 1 in \citet{Lee2000},
consistent with their work.
As can be seen, 
the model can account for the PV diagram reasonably well.

In the observations,
the power-law index of the mass-velocity relationship, $\gamma$, 
is calculated directly from the CO emission of the
molecular outflows.
Previously, $\gamma$ is known to be about 1.8 if
the emission is optically thin or if constant optical
depth emission was assumed for the CO emission
\citep{Masson1993,Cabrit1997}.
This $\gamma$ is consistent with the wind-driven model.
However, if the $^{12}$CO optical depth is corrected using
a velocity-dependent fit to the ratio $\tau(^{12}\textrm{CO}
J=1-0)/\tau(^{13}\textrm{CO}
J=1-0)$, then $\gamma$ is found to be between 2 and 4 (Yu 1999), which is
more consistent with the jet-driven model. 
Further observations are needed to
clarify the actual mass-velocity relationship.

\section{Conclusions}
We have performed a systematic study of the
basic properties of both
jet-driven and wind-driven molecular outflows using time-dependent
hydrodynamic simulations and simple analytic models. The main
conclusions are the following:
\begin{enumerate}
\item 
The structure and transverse velocity of the swept-up shell in 
a steady jet propagating into a uniform ambient medium 
can be reasonably reproduced by a ballistic bow shock model, presented in a
companion paper \citep{Ostriker2000}.
Since some material flows into the cocoon 
from the inner shell surface, and since mixing of \nam{}
material with  material already in the shell is not complete, the
ballistic bow shock model
only provides lower and upper limits for
the longitudinal velocity of the shell in the simulation.
\item
The PV diagrams along the outflow axis for 
the shell material swept-up by a steady jet 
show a convex spur structure with the highest velocity at the bow tip.
Low-velocity shell material is relatively symmetric in red and
bluesides at
any inclination; high velocity material is one-sided (either red or blue).
The power-law index of the mass-velocity relationship ranges
from 1.5 to 3.5, depending strongly on the inclination.
\item
In a pulsed jet, 
the internal bow shocks can affect the shell kinematics,
producing multiple convex structures in the PV diagram.
The internal bow shocks consist of jet material, and
have a higher forward velocity than the leading
bow shock, producing a high velocity gas signature in the PV diagrams
even at small inclination. The leading bow shock structure is not strongly
affected by internal bow shocks that collide with it obliquely.
\item 
The structure and kinematics of the shell in an isothermal steady
wide-angle wind simulation
can be well described by
a momentum-driven shell model similar to that of \citet{Shu1991}.
Since the thermal pressure is small compared to the
ram pressure of the wind, the shocked wind and ambient materials
merge into a thin shell of shocked material,
so that mass and momentum are conserved in each angular sector.
\item
In the steady wind simulation,
the PV diagrams cut along the outflow axis show a lobe 
structure tilted with inclination,
so that any given lobe is primarily red or blue except if the axis is
nearly in the plane of the sky.
Our wind models can not produce the
spur-like features seen in the jet simulations and a number of observed
systems. 
The power-law index of the mass-velocity
relationship ranges from 1.3 to 1.8.
\item
In a pulsed wind, 
the polar region of the internal surfaces is
flatter further out from the source and the wings of the internal surfaces 
interact with the shell producing multiple straight 
structures in the PV diagrams.
The internal surface consists of the wind material that
has a higher forward velocity than the leading
surface, producing high velocity gas signature in the PV diagrams
even at small inclination. 
\item The overall width of jet-driven shell is smaller than that of wind-driven
shell, even when the ambient medium
stratification is most favorable for narrow outflows in wind model and wide
outflows in jet model.
\end{enumerate}

There are clear differences between the 
PV diagrams and the power-law index of the mass-velocity relationship
between these two models. 
Comparing to observations, we find
that some outflows, e.g., HH 212, are consistent with the jet-driven model, 
while others, e.g., VLA 05487, are consistent with the wind-driven model.
Since the internal surface in
the wind model is much flatter than the observed H$_2$ bow shocks,
the temporal variation in the wind model
would need to occur within a small angle along the outflow axis to produce
such discrete curved structures,  and a significant decrease in the wind
velocity toward the equator would also be required.

 While none of our simple dynamical models of protostellar
outflows are consistent with {\it all} of
the detailed kinematics of individual sources,
this work shows that
broad similarities with characteristic signatures are easily identified.
The actual jet, wind,
and ambient material are undoubtedly
more complicated than our simple models,
due to the
intrinsic properties of the systems (e.g., magnetic fields, rotation,
density and velocity fluctuations in ambient gas).
Such complexities could potentially explain why some sources \cite[e.g. HH
111,][]{Lee2000} evidence signatures of both wind and jet driving in their
outflow kinematics.
It is possible that the differences
among observed outflows represent an evolutionary effect.
Further simulations are needed to investigate these and other intriguing
questions.

\acknowledgements
We thank an anonymous referee for a useful report.
We thank Mark McCaughrean
for providing the H$_2$ images of HH 212 and Peter Garnavich for VLA 05487.
This work was supported by DOE Grant DFG0398DP00215, NSF Grant 
AST-9981289 and NASA
Grant NAG-59575.

\clearpage


\clearpage

\figcaption{Distributions of number density, temperature and pressure
in logarithmic scale at 650 years in the steady jet simulation.
The vectors show the velocity structure. The solid line plotted over
the number density is the shell shape calculated from a 
ballistic bow shock model derived by \cite{Ostriker2000}. 
The gray-scale wedges along the right side indicate
the values for the gray-scale images. The z-axis is along the jet flow axis
and there is an assumed circular symmetry about the z-axis.
\label{fig:jetDTPV}}

\figcaption{Number density, temperature, and pressure 
along an axial cut through the head of the jet
in the steady jet simulation.
\label{fig:jetDTP}}

\figcaption{
Comparisons of the
velocity structures of the shell material
between the steady jet simulation and ballistic bow shock model.
Panels \tlabel{a} and \tlabel{b} show
the velocity structures for the simulation and model, respectively, 
in the bow shock frame plotted over the pressure distribution
of the simulation.
Panels \tlabel{c} and \tlabel{d} show the  
velocity structures for the simulation and model, respectively, 
in the observer's frame plotted over the density distribution 
of the simulation. 
In panels \tlabel{b} and \tlabel{d}, the open arrows are calculated with
the velocities of the \nam{} material and closed arrows are calculated
with the mean velocities of the shell material from the model.
\label{fig:jetP}}

\figcaption{Comparisons of the transverse velocity and longitudinal 
velocity of the shell material in the steady jet simulation (image)
with the ballistic bow shock model (lines).
Solid lines are calculated with the mean velocity of the shell material.
Dashed lines are calculated with the velocity of
the \nam{} material.
\label{fig:jetvrz}}

\figcaption{PV diagrams for the shell material 
at three inclinations cut along the outflow axis for the steady jet
simulation. $i$ is the inclination of the outflow to the plane of the sky.
Solid lines are calculated using the mean velocity of the shell material.
Dashed lines are calculated using the velocity of the \nam{}
material. Dotted lines indicate the zero velocity.
\label{fig:jetPV}}

\figcaption{Mass-velocity relationships at three
inclinations for the steady jet simulation. 
Both the redshifted (open square) and 
blueshifted (filled square) masses are shown.
The dashed lines are the fits to the redshifted mass with a power-law mass
velocity relationship, where the
power-law index, $\gamma$, is indicated at the upper 
right corner in each panel. 
The solid line at $i=0$\degree{} is calculated from the ballistic bow 
shock model.
\label{fig:jetmv}}

\figcaption{Distribution of number 
density in logarithmic scale at 610 years 
in the pulsed jet simulation,
with vectors showing the velocity structure.
PV diagrams and MV relationships at $i=0$\degree{}
and $i=30$\degree{} are also shown.
The gray-scale wedge along the right side indicates
the values for the number density.
Dashed lines in the MV relationships are fits 
to the redshifted mass with a power-law mass
velocity relationship, where the
power-law index, $\gamma$, is indicated at the upper
right corner.
\label{fig:pjetden}}

\figcaption{Distribution of number 
density in logarithmic scale at 390 years in the isothermal steady
wind simulation,
with vectors showing the velocity structure in the shell.
The solid line is the shell shape calculated from the
momentum-driven shell model. The gray-scale wedge
along the right side indicates
the values for the number density.
\label{fig:windiden}}

\figcaption{Transverse velocity, $v_R$, and longitudinal velocity, $v_z$,
of the shell material. The solid lines indicate
the velocity calculated from the momentum-driven shell model.
\label{fig:windivrz}}

\figcaption{PV diagrams for the shell material 
at three inclinations cut along the outflow axis.
$i$ is the inclination of the outflow to the plane of the sky.
The solid lines are the PV diagrams calculated
from the momentum-driven shell model.
\label{fig:windipv}}

\figcaption{Mass-velocity relationships
at three inclinations. 
Both the redshifted (open square) and 
blueshifted (filled square) masses are shown in the figure. 
The dashed lines are the fits to the redshifted mass with a power-law mass
velocity relationship, where the power-law index, $\gamma$, 
is indicated at the upper right corner in each panel.
The solid lines are calculated from the momentum-driven shell model.
\label{fig:windimv}}

\figcaption{Distributions of number density 
in logarithmic scale at 296 years in the isothermal pulsed wind
simulation. 
The vectors show the velocity structure in the shell and internal surfaces.
The gray-scale wedges along the right side indicate
the values for the gray-scale images.
PV diagrams and MV relationships at $i=0$\degree{} and $i=30$\degree{} 
are also shown.
Dashed lines in the MV relationships are fits 
to the redshifted mass with a power-law mass
velocity relationship, where the
power-law index, $\gamma$, is indicated at the upper
right corner.
\label{fig:pwindden}}

\figcaption{
CO emission, H$_2$ emission and PV diagram of the HH 212 outflow. 
On the left is the
CO emission contours \cite[][]{Lee2000}
plotted over the gray-scale
image of the H$_2$ emission \cite[][]{Zinnecker1998}. 
The outflow has been rotated clockwise by 23\degree{}, 
so that the outflow axis is N-S oriented.
On the right is the
PV diagram of the CO emission 
cut along the jet axis provided by \cite{Lee2000}.
Vertical dashed line indicates the ambient velocity around the outflow.
Horizontal dashed line indicates the position of the driving source of the
outflow. The solid lines plotted over the PV diagram are calculated with 
the ballistic bow shock model.
\label{fig:jetHH212}}

\figcaption{
CO emission, H$_2$ emission and PV diagram of the VLA 05487 outflow. 
On the left is the
CO emission contours \cite[][]{Lee2000}
plotted over the gray-scale
image of the H$_2$ emission \cite[][]{Garnavich1997}. 
The outflow has been rotated counterclockwise by 4\degree{}, 
so that the outflow axis is N-S oriented.
On the right is the
PV diagram of the CO emission 
cut along the jet axis provided by \cite{Lee2000}.
Vertical dashed line indicates the ambient velocity around the outflow.
Horizontal dashed line indicates the position of the driving source of the
outflow. The solid line plotted over the PV diagram is calculated with the 
modified X-wind model.
\label{fig:jetHH110}}

\setcounter{figure}{0}

%

\begin{figure}
\center
\epsscale{0.8}
\plotone{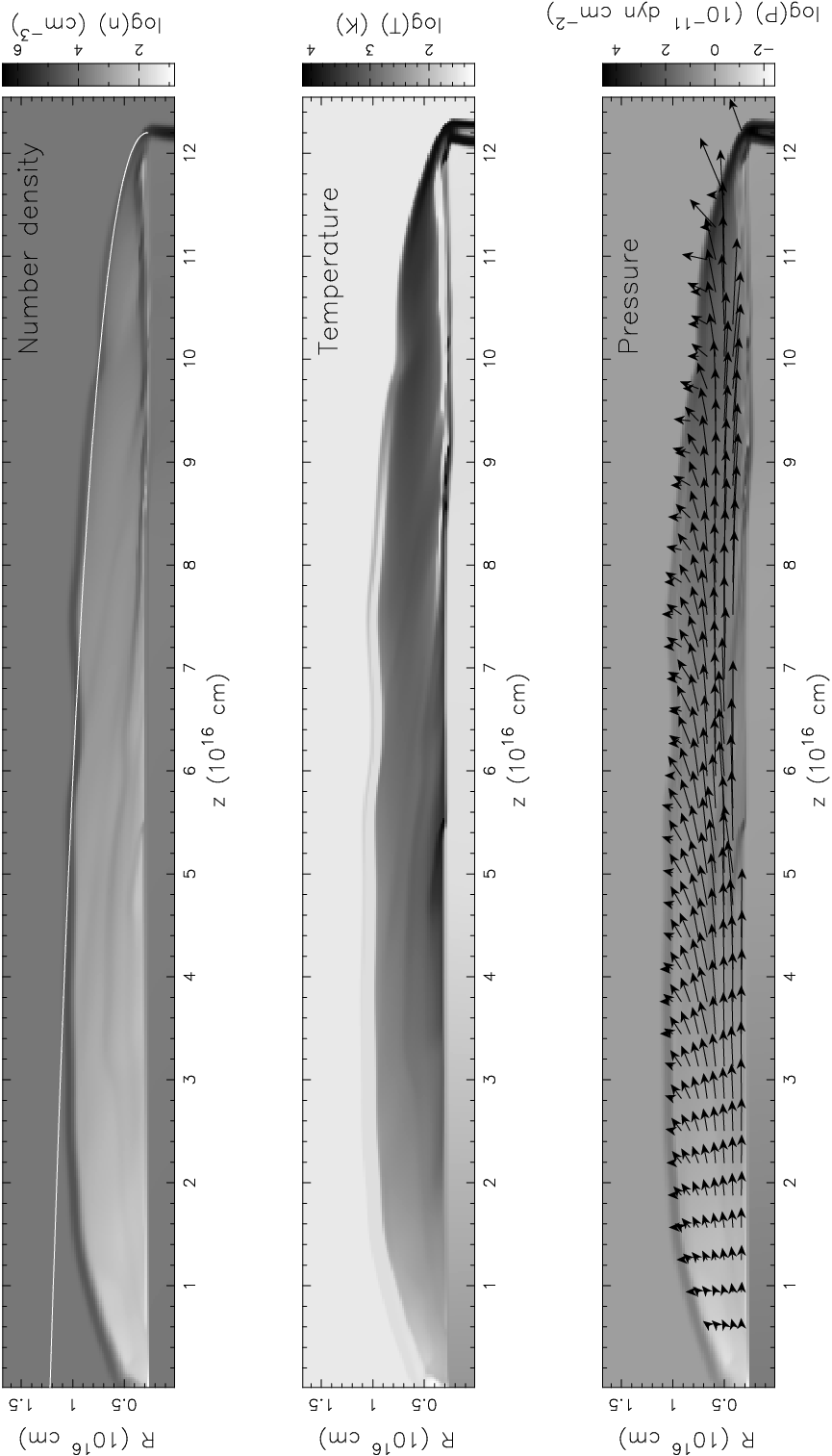}
\caption{}
\end{figure}

\clearpage
\begin{figure}
\center
\epsscale{0.8}
\plotone{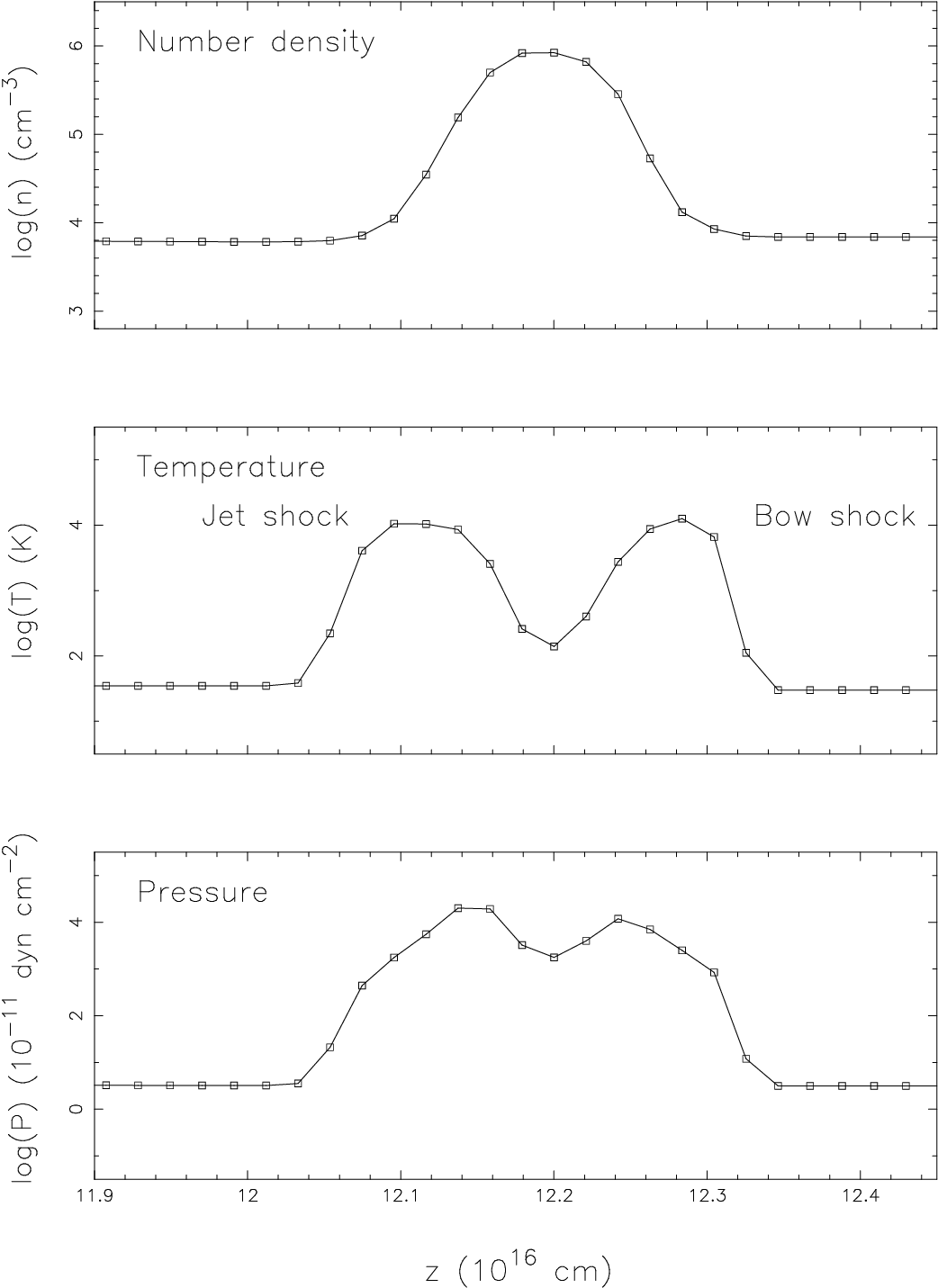}
\caption{}
\end{figure}

\clearpage

\begin{figure}
\center
\epsscale{0.85}
\plotone{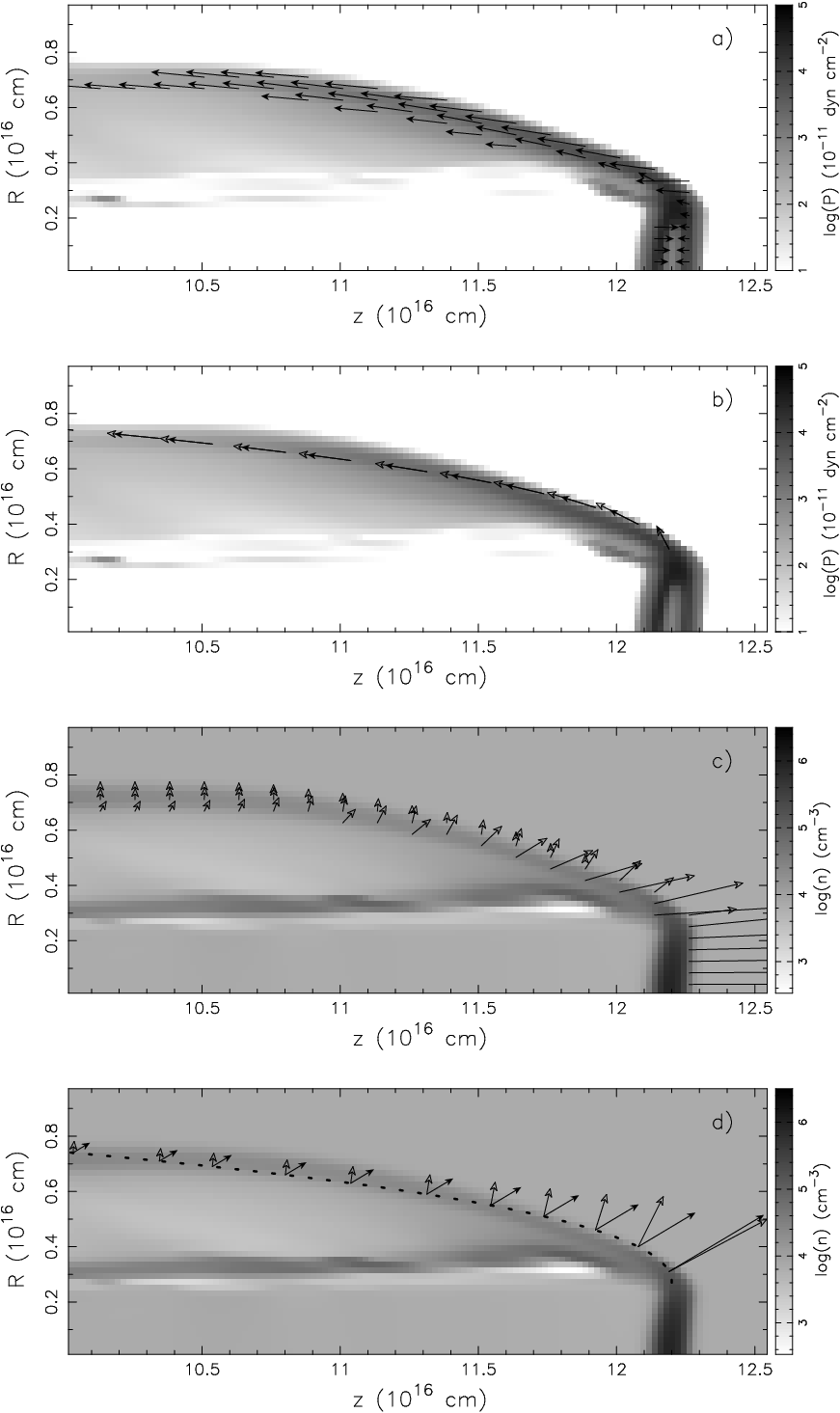}
\caption{}
\end{figure}

\clearpage
\begin{figure}
\center
\epsscale{1}
\plotone{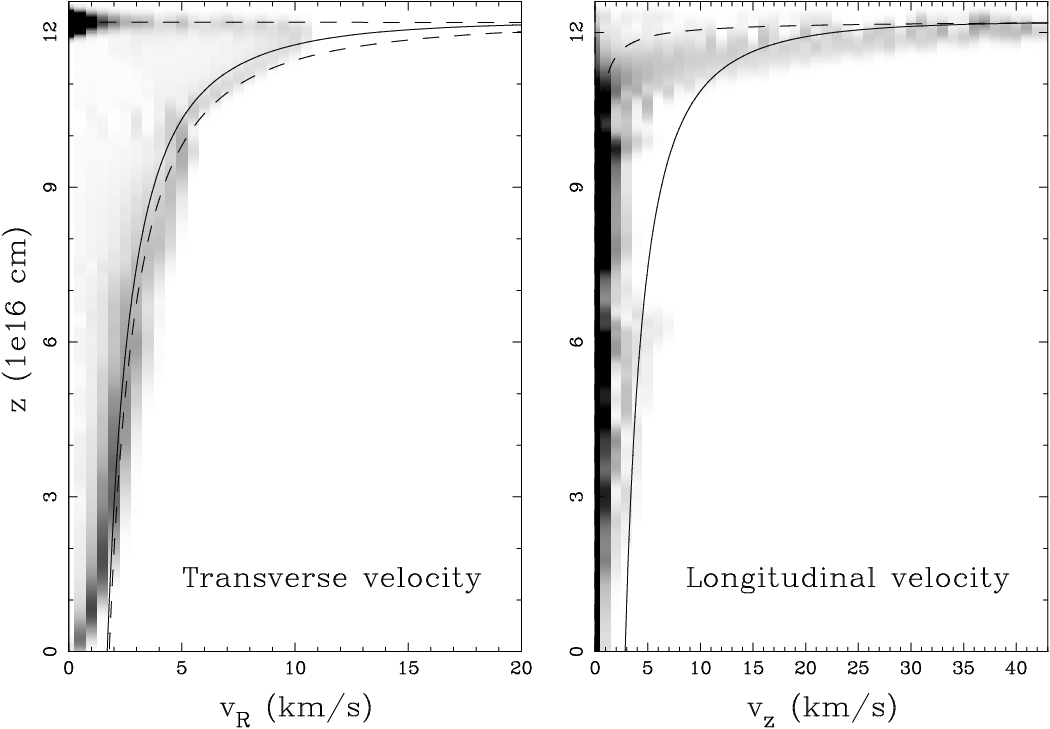}
\caption{}
\end{figure}

\clearpage

\begin{figure}
\center
\epsscale{0.8}
\plotone{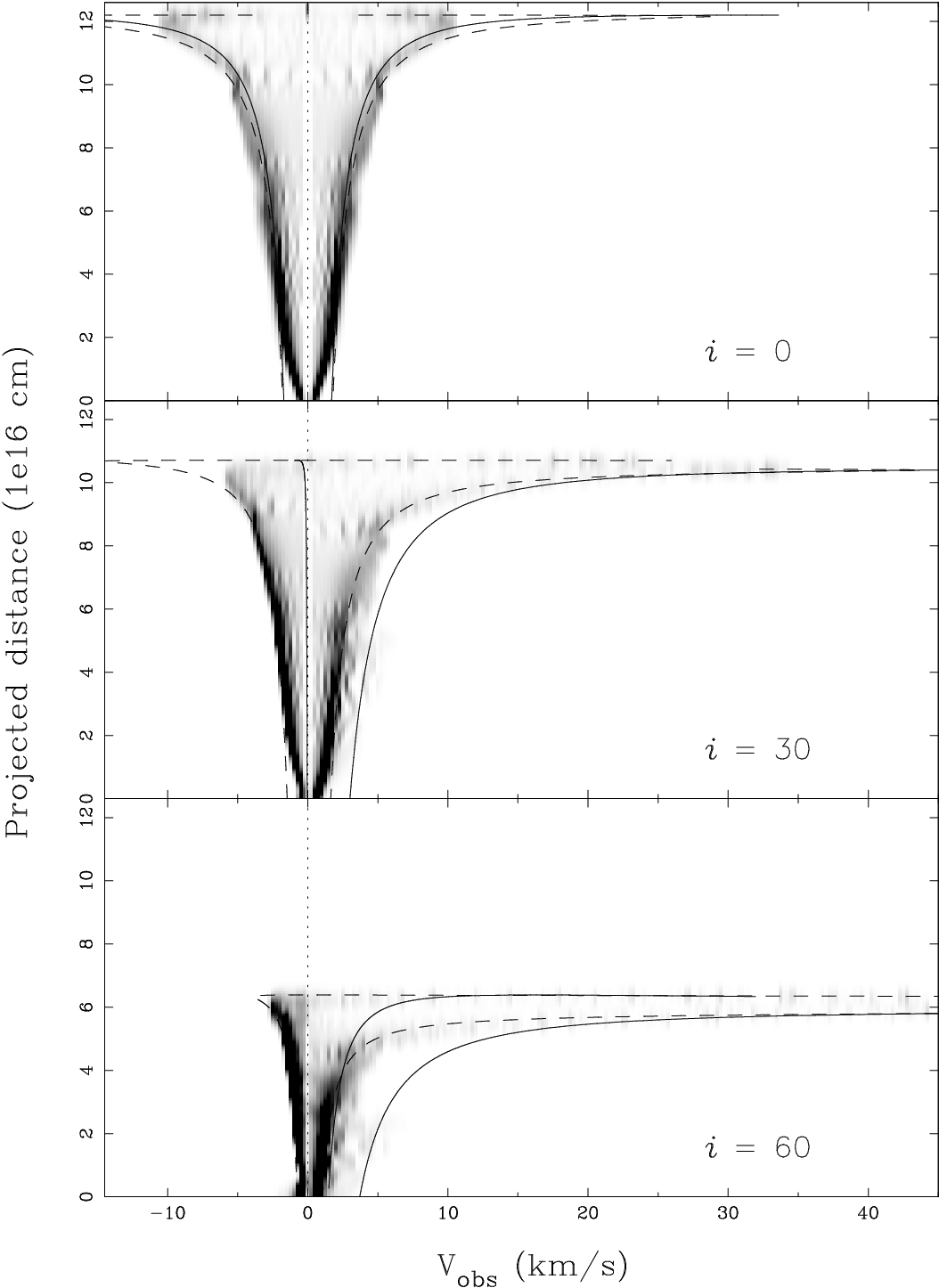}
\caption{}
\end{figure}

\clearpage

\begin{figure}
\center
\epsscale{0.8}
\plotone{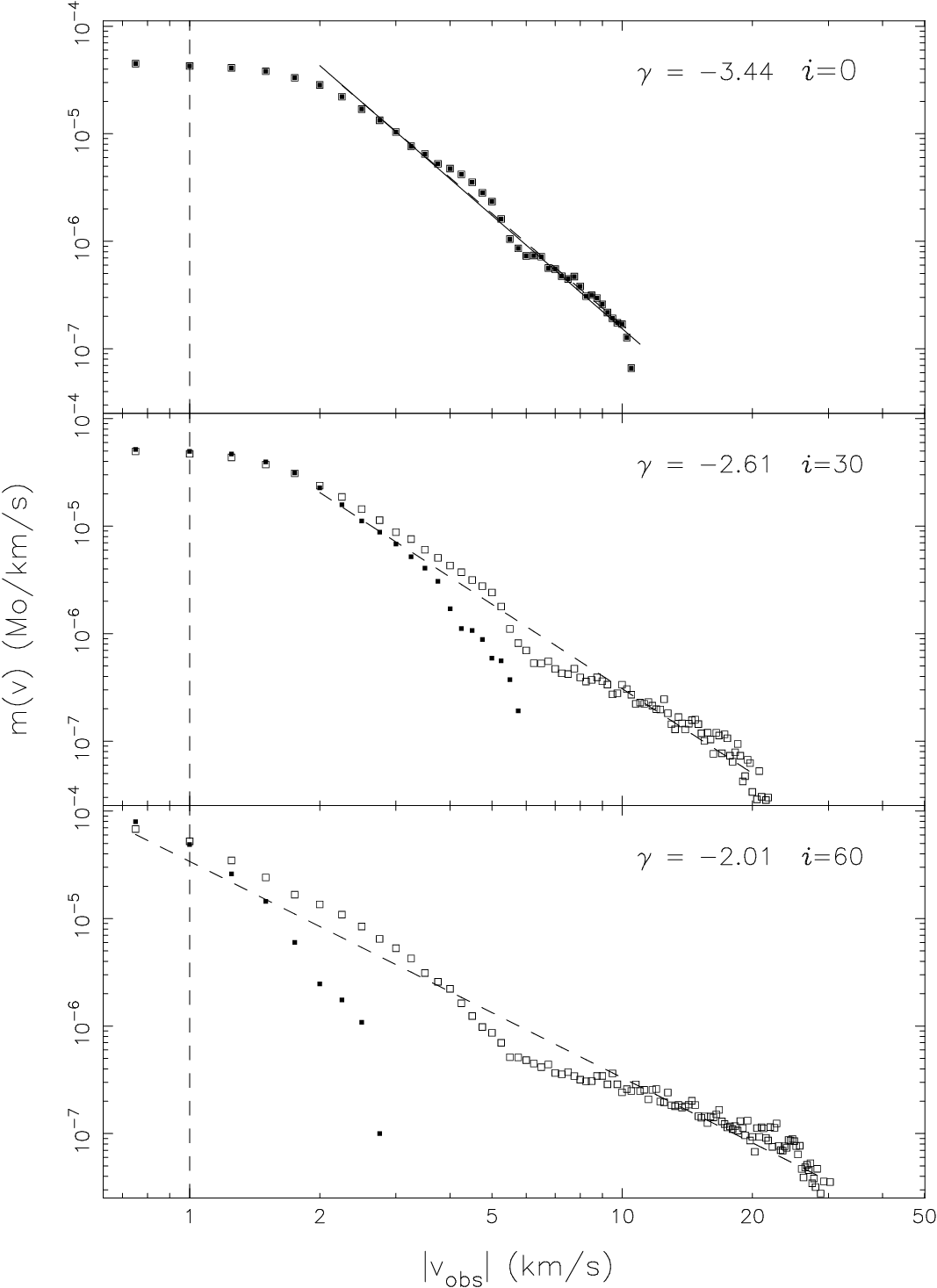}
\caption{}
\end{figure}

\clearpage

\begin{figure}
\center
\epsscale{0.8}
\plotone{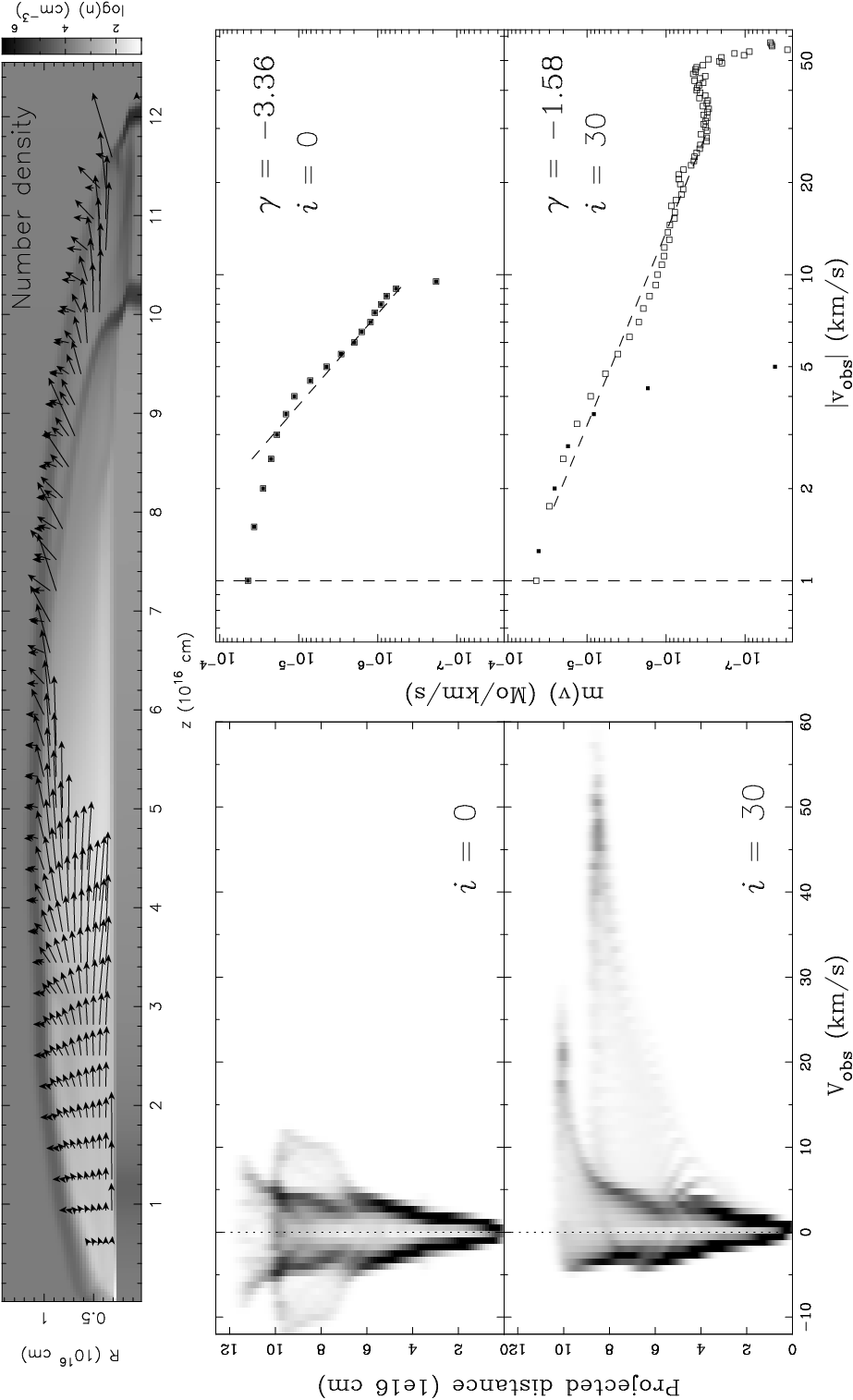}
\caption{}
\end{figure}

\clearpage

\begin{figure}
\center
\epsscale{0.25}
\plotone{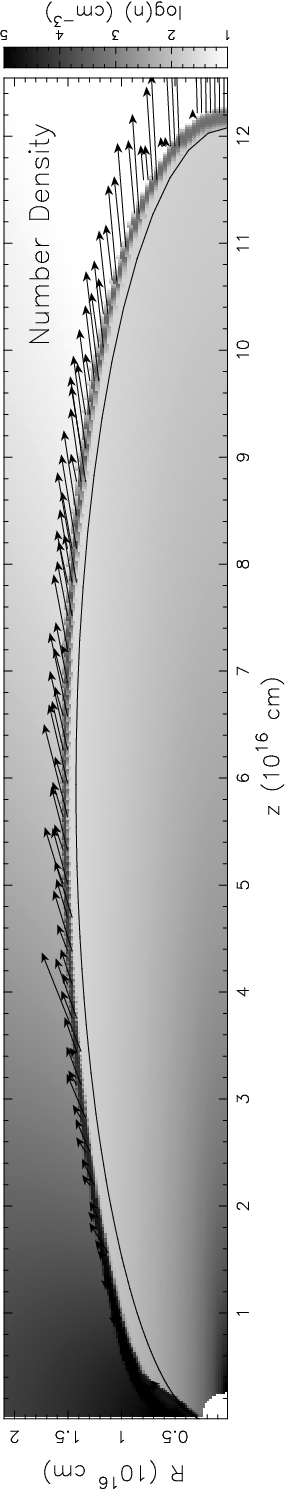}
\caption{}
\end{figure}

\clearpage

\begin{figure}
\center
\epsscale{1}
\plotone{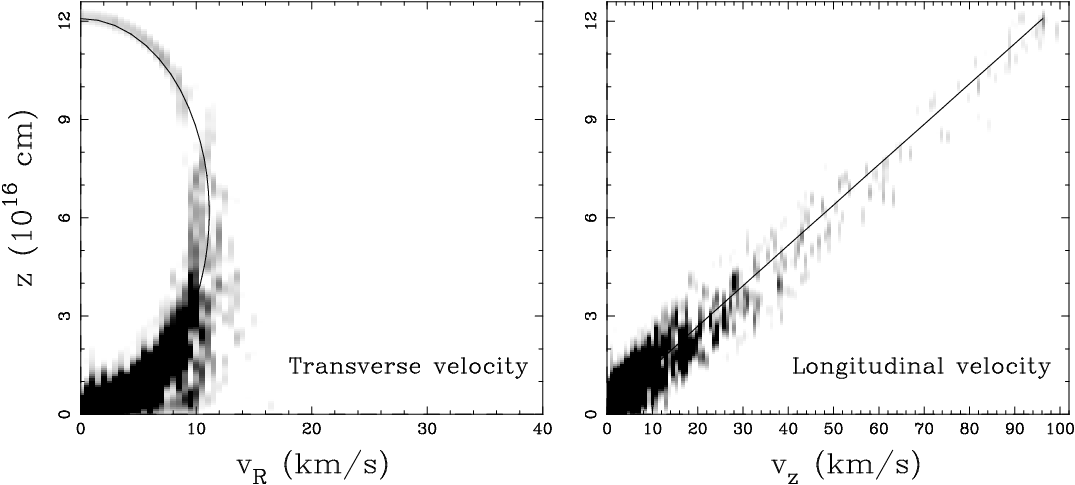}
\caption{}
\end{figure}

\clearpage

\begin{figure}
\center
\epsscale{0.8}
\plotone{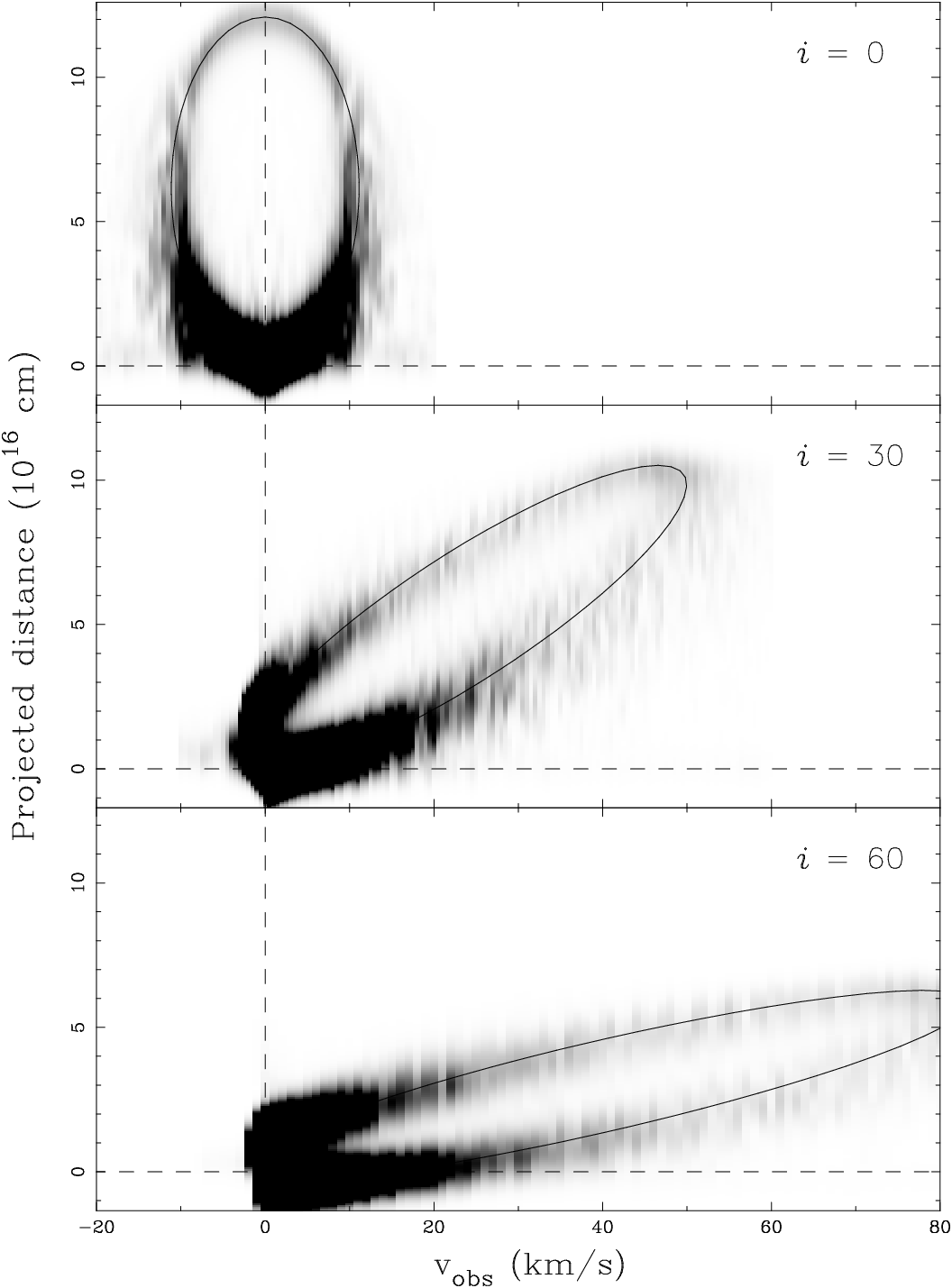}
\caption{}
\end{figure}

\clearpage

\begin{figure}
\center
\epsscale{0.8}
\plotone{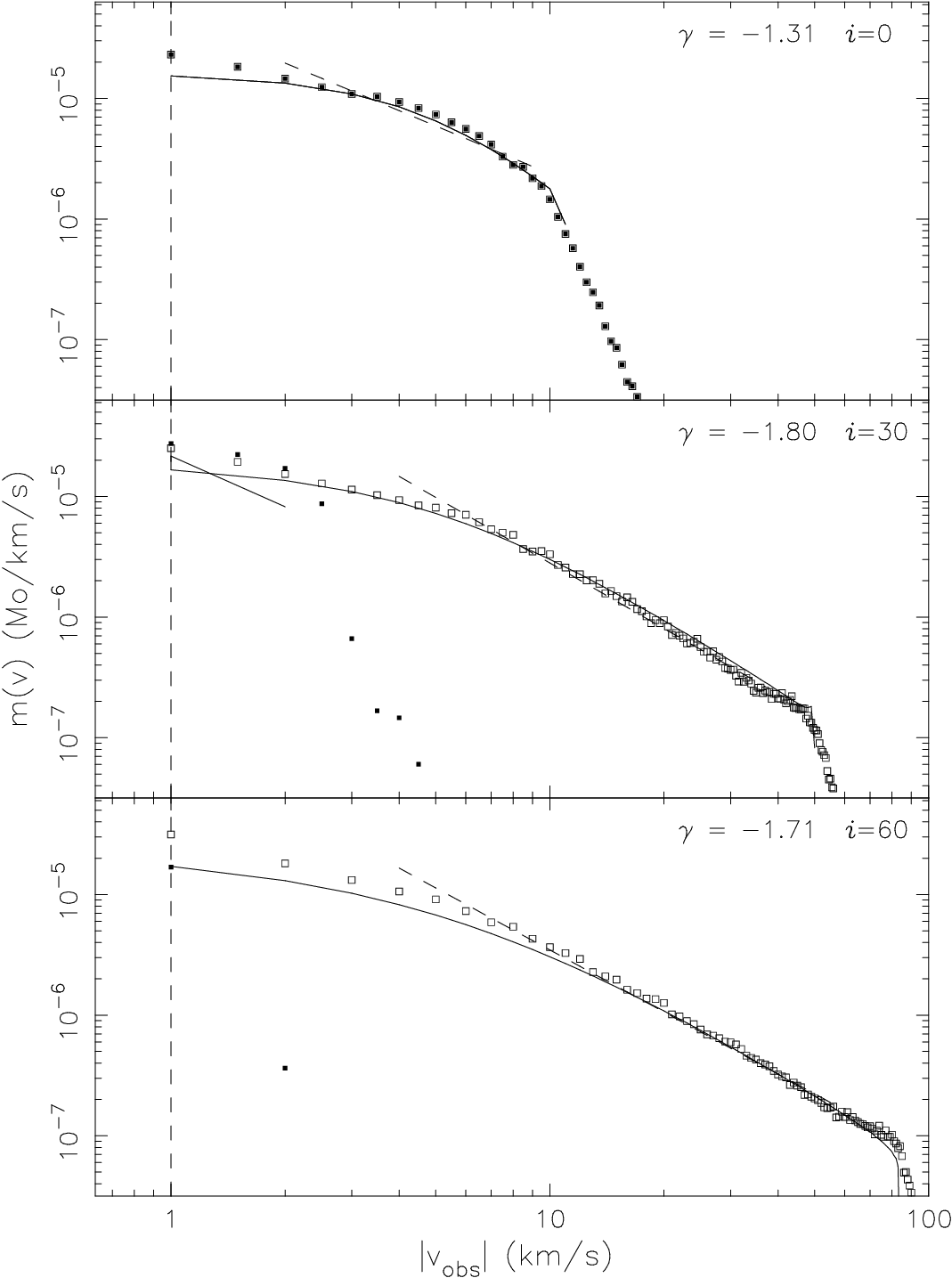}
\caption{}
\end{figure}

\clearpage

\begin{figure}
\center
\epsscale{0.8}
\plotone{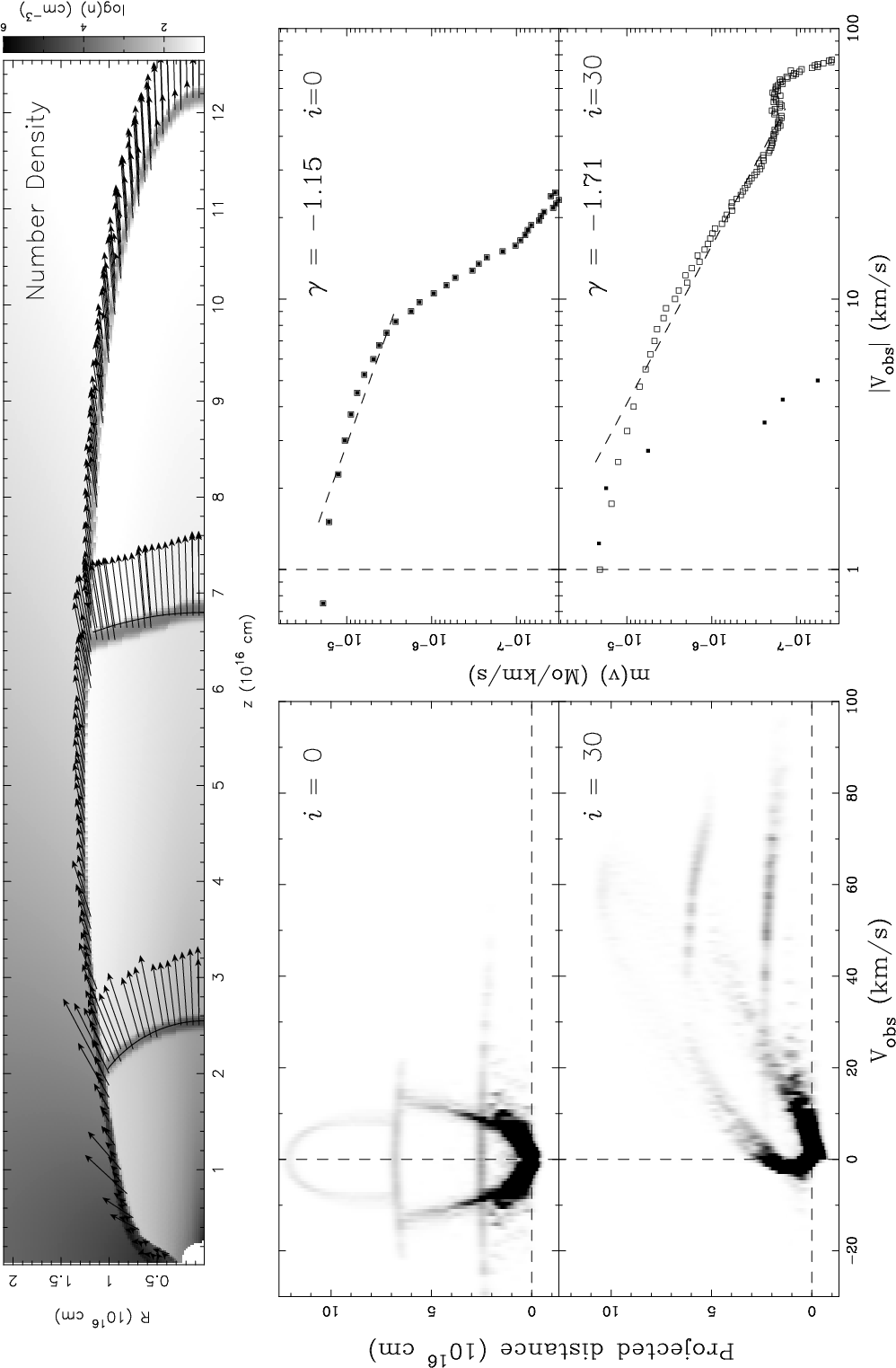}
\caption{}
\end{figure}

\clearpage

\begin{figure}
\center
\epsscale{0.7}
\plotone{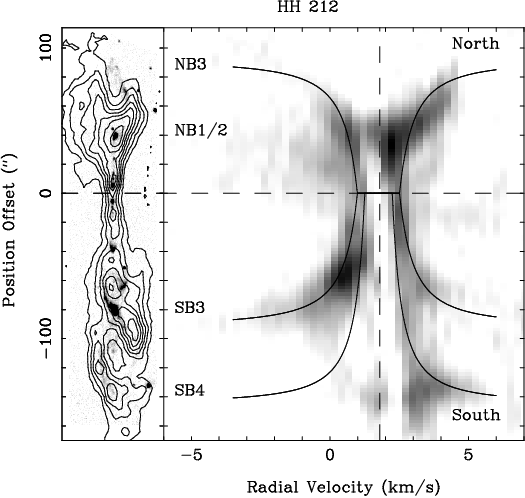}
\caption{}
\end{figure}

\clearpage

\begin{figure}
\center
\epsscale{0.7}
\plotone{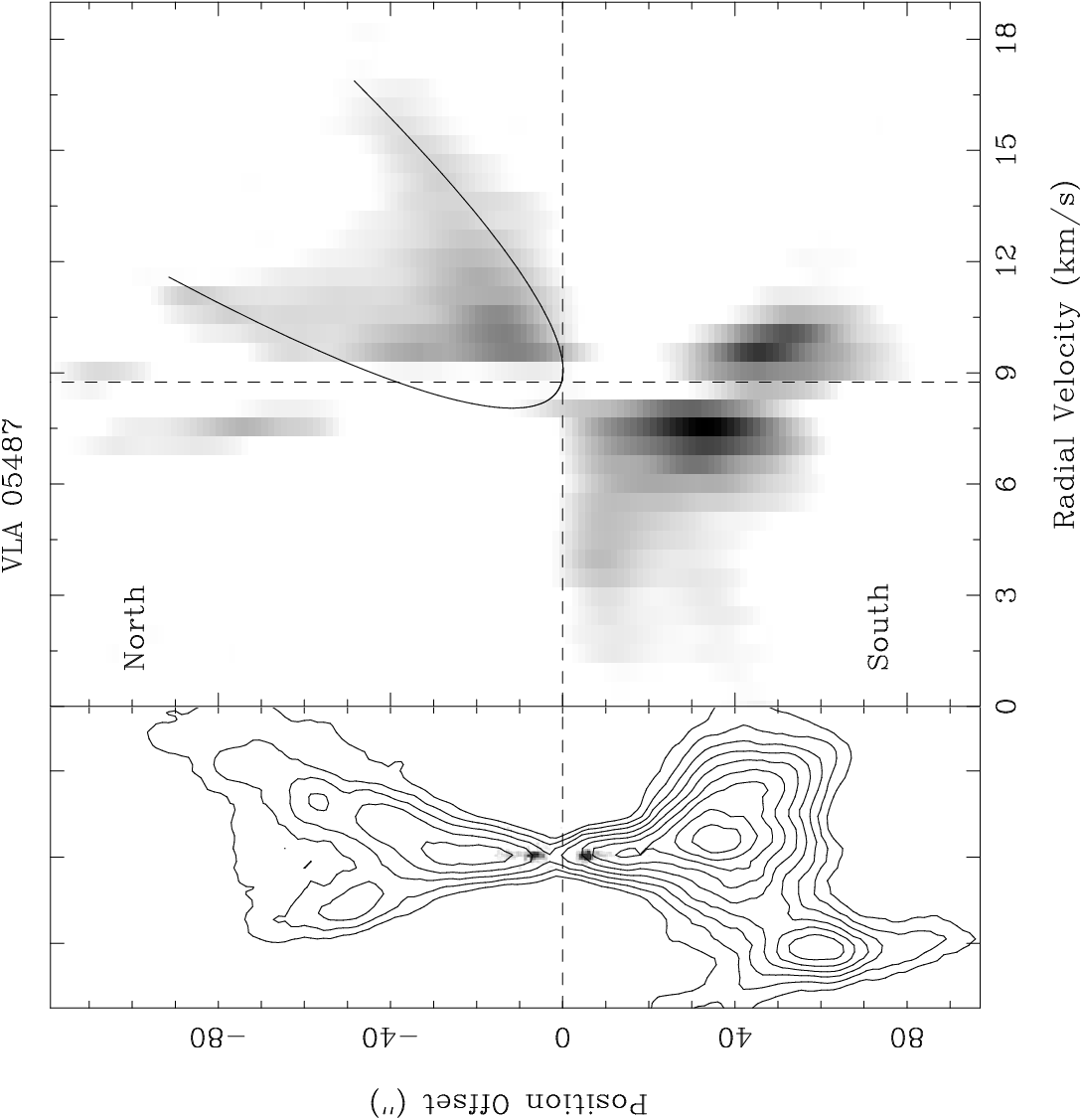}
\caption{}
\end{figure}

\clearpage


\begin{thebibliography}{}
\bibitem[Anglada(1995)]{Anglada1995} Anglada, G. 1995, Revista 
 Mexicana de Astronomia y Astrofisica Conference Series, 1, 67 
\bibitem[Bachiller(1996)]{Bachiller1996}Bachiller, R. 1996, ARAA, 34, 111
\bibitem[Bachiller et al.(1995)]{Bachiller1995} Bachiller, R., 
 Guilloteau, S., Dutrey, A., Planesas, P. \& Martin-Pintado, J. 1995, \aap, 
 299, 857 
\bibitem[Biro \& Raga(1994)]{Biro1994} Biro, S.\ and Raga, A.\ 
 C.\ 1994, \apj, 434, 221 
\bibitem[Blondin, Konigl \& Fryxell(1989)]{Blondin1989} Blondin, 
J.\ M., Konigl, A.\ and Fryxell, B.\ A.\ 1989, \apjl, 337, L37 
\bibitem[Blondin, Fryxell \& Konigl(1990)]{Blondin1990} Blondin, 
 J. M., Fryxell, B. A. \& Konigl, A.  1990, \apj, 360, 370 
\bibitem[Cabrit, Raga \& Gueth(1997)]{Cabrit1997} Cabrit, S., Raga, A., Gueth, F.
  1997, IAUS, 182, 163
\bibitem[Chernin et al.(1994)]{Chernin1994} Chernin, L.M., Masson, C.R.,
  Gouveia dal Pino, E.M., Benz, W., 1994, ApJ, 426, 204
\bibitem[Coppin, Davis, \& Micono(1998)]{Coppin1998} Coppin, K. E. 
  K., Davis, C. J. \& Micono, M.  1998, \mnras, 301, L10 
\bibitem[Dalgarno \& McCray(1972)]{Dalgarno1972} Dalgarno, A. \& 
 McCray, R. A. 1972, \araa, 10, 375 
\bibitem[Davis et al.(2000)]{Davis2000} Davis, C. J., 
Berndsen, A., Smith, M. D., Chrysostomou, A.\& Hobson, J. 2000, 
\mnras, 314, 241 
\bibitem[de Gouveia dal Pino \& Benz (1993)]{Gouveia1993} de Gouveia
dal Pino, E. M. \& Benz, W.  1993, \apj, 410, 686
\bibitem[Delamarter, Frank \& Hartmann(2000)]{Delamarter2000} 
 Delamarter, G., Frank, A. \& Hartmann, L. 2000, \apj, 530, 923 
\bibitem[Downes \& Ray(1999)]{Downes1999} Downes, T. P.  Ray, T. P. 
  1999, \aap, 345, 977 
\bibitem[Dutrey, Guilloteau \& Bachiller(1997)]{Dutrey1997} Dutrey, 
  A., Guilloteau, S. \& Bachiller, R. 1997, \aap, 325, 758 
\bibitem[Frank \& Mellema(1996)]{Frank1996} Frank, A.  \& Mellema, 
G.  1996, \apj, 472, 684 
\bibitem[Fukui et al.(1993)]{Fukui1993}Fukui, Y., Iwata, T., Mizuno, A., 
  Bally, J., Lane, A.P. 1993 in Protostars and Planets III, 
  ed. EH Levy, JI Lunine. Tucson; Univ. Ariz. Press
\bibitem[Garnavich et al.(1997)]{Garnavich1997} 
  Garnavich, P. M., Noriega-Crespo, A. , Raga, A. C. \& Bohm, K. -H.  1997,
  \apj, 490, 752
\bibitem[Gueth \& Guilloteau(1999)]{Gueth1999} Gueth, F. \& 
  Guilloteau, S. 1999, \aap, 343, 571 
\bibitem[Gueth, Guilloteau \& Bachiller(1996)]{Gueth1996} Gueth, 
  F., Guilloteau, S. \& Bachiller, R. 1996, \aap, 307, 891
\bibitem[Hollenbach(1997)]{Hollenbach1997} Hollenbach, D. 1997, IAU 
  Symp. 182: Herbig-Haro Flows and the Birth of Stars, 182, 181 
\bibitem[Hollenbach \& McKee(1979)]{Hollenbach1979} Hollenbach, D. \& 
  McKee, C. F. 1979, \apjs, 41, 555 
\bibitem[Lada(1985)]{Lada1985} Lada, C. J. 1985, \araa, 23, 267 
\bibitem[Lada \& Fich(1996)]{Lada1996} Lada, C. J. \& Fich, M.  
  1996, \apj, 459, 638 
\bibitem[Lee et al.(2000)]{Lee2000} Lee, C.-F., Mundy, L.G., Reipurth, B.,
  Ostriker, E.C., \& Stone, J.M. 2000, \apj, in press
\bibitem[Li \& Shu(1996a)]{Li1996a} Li, Z. -Y. \& Shu, F. H. 
1996a, \apj, 468, 261
\bibitem[Li \& Shu(1996b)]{Li1996b} Li, Z. -Y.  \& Shu, F. H. 
  1996b, \apj, 472, 211
\bibitem[MacDonald \& Bailey(1981)]{MacDonald1981} MacDonald, J. \& 
Bailey, M. E. 1981, \mnras, 197, 995 
\bibitem[Masson \& Chernin(1992)]{Masson1992} Masson, C. R. \& 
 Chernin, L. M. 1992, \apjl, 387, L47 
\bibitem[Masson \& Chernin(1993)]{Masson1993} Masson, C. R. \& 
  Chernin, L. M. 1993, \apj, 414, 230 
\bibitem[Matzner \& McKee(1999)]{Matzner1999} Matzner, C. D. \& 
  McKee, C. F. 1999, \apjl, 526, L109 
\bibitem[McKee et al.(1982)]{McKee1982} McKee, C.\ F., Storey, J.\ 
W.\ V., Watson, D.\ M.\ \& Green, S.\ 1982, \apj, 259, 647 
\bibitem[Mellema \& Frank(1997)]{Mellema1997} Mellema, G.  \& Frank, 
  A.  1997, \mnras, 292, 795 
\bibitem[Meyers-Rice \& Lada(1991)]{Meyers1991} Meyers-Rice, B. A. 
 \& Lada, C. J. 1991, \apj, 368, 445 
\bibitem[Micono et al.(1998)]{Micono1998} Micono, M., Davis, C. J., 
  Ray, T. P., Eisloeffel, J. \& Shetrone, M. D. 1998, \apjl, 494, L227 
\bibitem[Nagar et al.(1997)]{Nagar1997} Nagar, N. 
 M., Vogel, S. N., Stone, J. M. \& Ostriker, E. C. 1997, \apjl, 482, L195
\bibitem[Najita \& Shu(1994)]{Najita1994} Najita, J.\ R.\ \& Shu, 
F.\ H.\ 1994, \apj, 429, 808 
\bibitem[Ostriker(1997)]{Ostriker1997} Ostriker, E. C. 1997, \apj, 486, 291
\bibitem[Ostriker(1998)]{Ostriker1998} Ostriker, E. C. 1998, 
  Accretion Processes in Astrophysical Systems: Some Like it Hot!, 484 
\bibitem[Ostriker et al.(2000)]{Ostriker2000} {Ostriker, E. C., Lee, C.-F., 
 Stone, J.M. and Mundy, L.G. 2000, \apj, submitted}
\bibitem[Raga \& Cabrit(1993)]{Raga1993} Raga, A. \& Cabrit, S. 
 1993, \aap, 278, 267
\bibitem[Raga et al.(1990)]{Raga1990} Raga, 
  A. C., Binette, L. , Canto, J.  \& Calvet, N.  1990, \apj, 364, 601  
\bibitem[Ray, et al.(1996)]{Ray1996} Ray, T. P., Mundt, R. , 
  Dyson, J. E., Falle, S. A. E. G. \& Raga, A. C. 1996, \apjl, 468, L103 
\bibitem[Reipurth \& Graham(1988)]{Reipurth1988} Reipurth, B. \& 
 Graham, J. A. 1988, \aap, 202, 219 
\bibitem[Reipurth \& Olberg(1991)]{Reipurth1991} Reipurth, B. \& 
  Olberg, M.  1991, \aap, 246, 535 
\bibitem[Reipurth, Bally, \& Devine(1997)]{Reipurth1997} Reipurth, B., 
 Bally, J.  \& Devine, D.  1997, \aj, 114, 2708 
\bibitem[Reipurth et al.(1999)]{Reipurth1999} Reipurth, B., Yu, K. , 
  Rodriguez, L. F., Heathcote, S.  \& Bally, J.  1999, \aap, 352, L83 
\bibitem[Richer et al.(2000)]{Richer2000}Richer, J. Shepherd, D., 
  Cabrit, S., Bachiller, R., \& Churchwell, E. 2000, in Protostars
  and Planets IV, ed. V. Mannings, A. P. Boss \& S. S. Russell
  (Tucson: University of Arizona Press), in press
\bibitem[Shang et al.(1998)]{Shang1998} Shang, H. , Shu, 
  F. H. \& Glassgold, A. E. 1998, \apjl, 493, L91 
\bibitem[Shu et al.(1991)]{Shu1991} Shu, F. H., 
  Ruden, S. P., Lada, C. J. \& Lizano, S.  1991, \apjl, 370, L31 
\bibitem[Shu et al.(1994)]{Shu1994} Shu, F.\ H., Najita, J., Ruden, 
S.\ P., \& Lizano, S.\ 1994, \apj, 429, 797 
\bibitem[Shu et al.(1995) 1995]{Shu1995} Shu, F. 
  H., Najita, J. , Ostriker, E. C. \& Shang, H.  1995, \apjl, 455, L155  
\bibitem[Shu et al.(2000)]{Shu2000} Shu, F.H.,  Najita, J., Shang, H., 
  \& Li, Z. -Y. 2000, in Protostars
  and Planets IV, ed. V. Mannings, A. P. Boss \& S. S. Russell
  (Tucson: University of Arizona Press), in press
\bibitem[Smith, Suttner, \& Yorke(1997)]{Smith1997} Smith, M. D., 
  Suttner, G. \& Yorke, H. W. 1997, \aap, 323, 223 
\bibitem[Stahler(1994)]{Stahler1994} Stahler, S. W. 1994, \apj, 
 422, 616
\bibitem[Stone \& Norman(1992)]{Stone1992} Stone, J. M. \& Norman, 
 M. L. 1992, \apjs, 80, 753
\bibitem[Stone \& Norman(1993a)]{Stone1993a} Stone, J. M. \& Norman, 
 M. L. 1993, \apj, 413, 198  
\bibitem[Stone \& Norman(1993b)]{Stone1993b} Stone, J.\ M.\ and 
Norman, M.\ L.\ 1993, \apj, 413, 210 
\bibitem[Stone \& Norman(1994)]{Stone1994} Stone, J. M. \& Norman, 
 M. L. 1994, \apj, 420, 237 
\bibitem[Suttner et al.(1997)]{Suttner1997} 
  Suttner, G., Smith, M. D., Yorke, H. W. \& Zinnecker, H. 1997, \aap, 318, 
  595 
\bibitem[\Volker{} et al.(1999)]{Volker1999}
  \Volker, R. , Smith, M. D., Suttner, G.  \& Yorke, H. W. 1999, 
  \aap, 343, 953 
\bibitem[Yu, Billawala \& Bally(1999)]{Yu1999} Yu, K. , 
  Billawala, Y.  \& Bally, J.  1999, \aj, 118, 2940 
\bibitem[Zhang \& Zheng(1997)]{Zhang1997} Zhang, Q.  \& Zheng, X.  
  1997, \apj, 474, 719 
\bibitem[Zinnecker, McCaughrean, \& Rayner(1998)]{Zinnecker1998} 
  Zinnecker, H., McCaughrean, M. J. \& Rayner, J. T. 1998, \nat, 394, 862 
\end{thebibliography}
\end{document}